\documentclass{article}

\usepackage{arxiv}

\usepackage[utf8]{inputenc} 
\usepackage[T1]{fontenc}    
\usepackage{hyperref}       
\usepackage{url}            
\usepackage{booktabs}       
\usepackage{amsfonts, amsthm,amsmath,amssymb}       
\usepackage{nicefrac}       
\usepackage{microtype}      
\usepackage{lipsum}		
\usepackage{graphicx}
\usepackage{natbib}
\usepackage{doi}

\usepackage{amsthm, bm}
\usepackage{multirow}
\usepackage{booktabs} 
\newcommand\independent{\protect\mathpalette{\protect\independenT}{\perp}}
\def\independenT#1#2{\mathrel{\rlap{$#1#2$}\mkern2mu{#1#2}}}
\usepackage{subcaption}
\usepackage[ruled, english]{algorithm2e}
\renewcommand{\algorithmcfname}{Algorithm}%
\SetKwInput{KwInput}{Input}%
\SetKwInput{KwOutput}{Output}%
\SetKwInput{KwData}{Data}%
\SetKwInput{KwResult}{Result}%
\SetKwFor{For}{for}{do}{end}%
\SetKwFor{While}{while}{do}{end}%

\usepackage{geometry}
\theoremstyle{plain}

\newtheorem{theorem}{Theorem}[section]

\theoremstyle{remark}
\newtheorem{definition}[theorem]{Definition}

\title{A proposal of smooth interpolation to optimal transport for restoring biased data for algorithmic fairness}

\date{} 					

\author{ \href{https://orcid.org/0009-0000-2966-3404}{\includegraphics[scale=0.06]{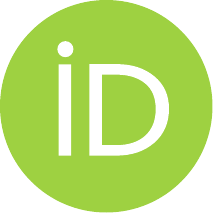}\hspace{1mm}Elena M.~De-Diego}\\
	Institute of Data Science and Artificial Intelligence (DATAI), \\
	Universidad de Navarra, Ismael Sánchez Bella Building, Campus Universitario\\
	31009 Pamplona, Spain \\
	\texttt{e.dediego.m@gmail.es} \\
	\And
	\href{https://orcid.org/0000-0002-0455-1200}{\includegraphics[scale=0.06]{orcid.pdf}\hspace{1mm}Paula~Gordaliza} \\
	Institute for Advanced Materials and Mathematics (INAMAT$^2$)\\
	Universidad Pública de Navarra\\
	31009 Pamplona, Spain \\
	\texttt{paula.gordaliza@unavarra.es} \\
	\And
	\href{https://orcid.org/0000-0001-7502-8188}{\includegraphics[scale=0.06]{orcid.pdf}\hspace{1mm}Jesús~López-Fidalgo} \\
		Institute of Data Science and Artificial Intelligence (DATAI), \\
	Universidad de Navarra, Ismael Sánchez Bella Building, Campus Universitario\\
	31009 Pamplona, Spain \\
	\texttt{fidalgo@unav.es} \\
}

\renewcommand{\headeright}{Under review}
\renewcommand{\undertitle}{Under review}
\renewcommand{\shorttitle}{PLS-based approach for fair representation learning}

\hypersetup{
	pdftitle={A proposal of smooth interpolation to optimal transport for restoring biased data for algorithmic fairness},
	pdfsubject={Statistics, Applications},
	pdfauthor={Paula Gordaliza},
	pdfkeywords={Algorithmic fairness, Demographic parity, Optimal transport, Smooth interpolation, Cyclical monotonicity, Distribution-freeness},
}

\begin{document}
	\maketitle
	
\begin{abstract}
The so-called algorithmic bias is a hot topic in the decision making process based on Artificial Intelligence, especially when demographics, such as gender, age or ethnic origin, come into play. 
Frequently, the problem is not only in the algorithm itself, but also in the biased data feeding the algorithm, which is just the reflection of the societal bias. 
Thus, this `food' given to the algorithm have to be repaired in order to produce unbiased results for all. 
As a simple, but frequent case, two different subgroups will be considered, the privileged and the unprivileged groups. Assuming that results should not depend on such characteristic dividing the data, the rest of attributes in each group have to be moved (transported) so that its underlying distribution can be considered similar. 
For doing this, optimal transport (OT) theory is used to effectively transport the values of the features, excluding the sensitive variable, to the so-called {\it Wasserstein barycenter} of the two distributions conditional to each group. An efficient procedure based on the {\it auction algorithm} is adapted for doing so. 
The transportation is made for the data at hand. If new data arrive then the OT problem has to be solved for the new set gathering previous and incoming data, which is rather inefficient. 
Alternatively, an implementation of a smooth interpolation procedure called \textit{Extended Total Repair (ExTR)}  is proposed, which is one of the main contributions of the paper. 
The methodology is applied with satisfactory results to simulated biased data as well as to a real case for a German credit dataset used for risk assessment prediction.
\end{abstract}



\section{Introduction}
\label{sec:introduction}
{\it Machine learning} (ML)-based systems have garnered considerable significance in the context of data-driven decision-making, increasingly used in a large number of sensitive domains, including healthcare \citep{Morik10}, finance \citep{10.5555/573193}, and criminal justice \citep{angwin2016machine, DBLP:journals/corr/abs-1905-12728}. 
Therefore, the consequential impact of automated predictions, built under statistical models, directly affects human lives in many ways. Indeed, several works have proven that data-driven algorithms may inherit, reproduce, or even amplify existing patterns within the historical data, with which they are trained, related to societal prejudices showing inequalities between different demographic groups of age \citep{CUI2022106029}, gender \citep{10.5555/3157382.3157584} or ethnicity \citep{pmlr-v81-buolamwini18a}, for instance. 
In general, certain variables within the dataset, referred to as {\it protected variables}, may entail sensitive information that needs to be carefully processed to ensure fair and equitable predictions. 
At the intersection between {\it Artificial Intelligence} (AI) and ethics, fair learning has established itself as a very active area of research in the last decade, whose purpose is to ensure that predictive algorithms are not discriminatory towards any individual or subgroup of population, based on such protected characteristics.
For a recent survey on this topic, we refer to \cite{10.1145/3616865} or \cite{10.1145/3597199}.
\par
In particular, within the context of fair supervised learning, a decision rule is trained to learn the relationship between the input $X \in \mathcal{X}$  (unprotected variables) and the output $Y \in \mathcal{Y}$, in presence of sensitive information contained in the protected variables $S \in \mathcal{S}$. It is important to clarify three issues at the outset. First of all, we note that we consider a bias-aware learning framework, where $S$ is known and observed, therefore it is not explicitly used to learn $g$ \citep{10.1145/2090236.2090255}. This does not mean that sensitive information will not be learned due to the correlations between the input variables \citep{Besse2022}. Secondly, the interest here relies in achieving group fairness, which emphasizes an equal treatment of different groups based on sensitive features $S$ \citep{10.1145/3351095.3372864, 10.5555/3060832.3061001}. Third, in this work we focus on fair classification, where the objective is then to find the best possible rule $g: \mathcal{X} \rightarrow \mathcal{Y}$ by minimizing the empirical risk computed on a finite training set of labeled samples i.i.d. from $(X,Y)$, where $x\in \mathcal{X} \subset \mathbb{R}^{d}, \ d\geq 1$ and $Y \in \mathcal{Y} = \{0,1\}$; while ensuring that $S$ has no influence on the outcome $\hat{Y}=g(X)$. 
\par
From a mathematical point of view, the underlying idea of group fairness can be expressed in terms of the statistical independence between the variables involved in the learning process. In this sense, the independence $\hat{Y}\independent S$ is called
{\it Demographic Parity} (DP) \citep{10.1145/2783258.2783311}; whereas when the true value of the target $Y$ is known, the conditional independence $\hat{Y}\independent S | Y$ is known as {\it Equality of Odds} (EO) \citep{NIPS2016_9d268236}.
We refer to \cite{} for a detailed study of other fairness definitions.
In order to measure fairness in sets other than the training set, for which information about $Y$ is probably not available, we will focus on DP. Moreover we consider a binary protected variable $S \in \mathcal{S}=\{0,1\}$, meaning that the population can be divided into two groups $S = 0$, for the unfavored class, and $S = 1$, for the favored class. In this setting, a classifier $g: \mathbb{R}^{d} \rightarrow \{0, 1\}$ is said to achieve DP, with respect to the joint distribution of $(X, S)$, if $P(g(X) = 1 | S = 0) = P(g(X) = 1 | S = 1)$, which is usually quantified through 
the {\it disparate impact} (DI) of the classifier $g$
as $DI(g, X, S) = \frac{P(g(X) = 1 | S = 0)}{P(g(X) = 1 | S = 1)}$ that measures the level of DP attained. Note that $DI(g,X,S)=1$ characterizes a fully fair rule (in the sense of DP, which will not be mentioned hereinafter). It is worth to mention that these measures are based on prediction made with a classifier. This is actually the  objective, a fair classification. Nevertheless, in this paper we will repair the data feeding the algorithm for constructing the classifier. Thus, this work is valid and useful for any  offering fair data for the training.
\par
Whichever notion of fairness is chosen by practitioners, methods for improving algorithmic fairness can be divided into pre-processing, in-processing or post-processing, depending on the time when such fairness conditions are imposed. We refer to \cite{Barocas2018FairnessAM, Oneto_2020} for comprehensive studies on fairness enhancing methods and the references therein. Generally, a realistic scenario in practice assumes that the algorithm is unknown and inaccessible, therefore not modifiable. Moreover, as observed by \cite{}, the choice of the learning set holds significant influence over the resulting rule. Therefore, we focus on the potential strategy to enhance fairness by pre-processing the training data to remove any possible dependencies with the sensitive information. This has been mainly done in the literature through fair representation learning, which involves mapping the training data to a latent space where the dependencies between protected attribute and class labels disappear, while preserving the accuracy \citep{10.1145/2783258.2783311, johndrow2017algorithm}. To this end, one of the most satisfactory approaches is provided by optimal transport (OT) theory.
\par
In recent years, thanks to the development of powerful computational methods, OT is gaining increasing interest in many different domains, particularly in ML, where it is essential to compare probability distributions. Appendix \ref{app:sec:background:OT} briefly describes the fundamentals of OT and mentions further application work in different ML problems. In the fair learning framework, OT techniques were first applied to the problem of fair binary classification by \cite{10.1145/2783258.2783311} from a computational point of view, and then formalized mathematically for quadratic transport cost by \cite{pmlr-v97-gordaliza19a}. In particular, authors established links between predictability (a particular version of the classification error), the DI of a classifier, and the total variation distance between the probability distributions of the two demographic groups, namely $\mu_s:=\mathcal{L}(X | S = s), \ s = 0, 1$. Importantly, they provided an upper bound for the minimal excess risk in terms of the Wasserstein variation (of order $2$) between the distributions $\mu_s$ that support the use of their weighted Wasserstein barycenter $\mu_B$ (with weights $\pi_{s} := P(S = s)$) as the distribution of a fair representation of the original input data $X$, which is referred to as a \textit{repair} of the data. Later, other approaches \citep{jiang2020wasserstein, silvia2020general, chzhen2020fair, buyl2022optimal} proposed to apply OT to modify predictions.
\par
In the practical setup, the repair methodology given by \cite{pmlr-v97-gordaliza19a} consists in computing the empirical Wasserstein barycenter between the empirical counterparts of $\mu_S$, which is reduced to obtaining the discrete optimal transport map between them. However, this proposal imposes a clear limitation due to the fact that such a map is derived from the given data, and therefore defined on a specific discrete support.  Alternatively, a continuous extension of the OT map can be employed to handle samples that fall outside the support of the empirical distributions of $\mu_S$.
This problem arises, for example, in any online system that receives streamed data. There,
the naive approach would be to recalculate the transport plan between the augmented subsets, yet incurring in a significant computational cost. In addition, another clear application is in an algorithmic evaluation process by means of K-fold cross-validation (CV), where different sets of train and test sets are handled. In this situation, 
the repair would be done for the training set in each split, and then the interpolation can be used to repair the data in the test set. Although the repair (and the interpolation) could actually be done only once,
we did it here for all splits just to show the robustness of the procedure.
\par
The first proposal introduced by \cite{cuturi2013sinkhorn} to efficiently obtain a differentiable approximation to empirical OT  consists in adding an entropic regularization term to the optimization problem. Since then, given the potential of this tool, this problem has received much attention and great advances  on the rates of convergence have been obtained in the last decade. Recently, \cite{vacher2021dimension} proved that the squared Wasserstein distance
between two distributions could be approximately computed in polynomial time with appealing statistical error bounds. Later \cite{vacher2024optimal} extended this result  to  the problem of estimating  the transport map between two distributions in $L^2$ distance. We refer also to these two works for a brief review of this approach.
\par
Alternatively, inspired  by McCann's main theorem  \citep[p~310]{10.1215/S0012-7094-95-08013-2}, another different direction, based on ranks and signs, to obtain an smooth interpolation of OT maps was explored by \cite{10.1214/20-AOS1996} as a necessary tool for extending the univariate concepts of distribution and quantile functions to the multivariate context. In this work, the convergence of the interpolation to the true OT map is proven when $\mu_1$ is the spherical uniform law over the $d-$dimensional unit open ball.
Later \cite{delara2021consistent}  extended the consistency of such estimator to more general spaces.
Furthermore, the authors applied it to develop empirically-based counterfactual explanations, extending the work of \cite{wachter2018counterfactual}. 
Nevertheless, their work is limited to theoretical results, without a proposal to implement it.
\par
In this work, we follow the guidelines of the second proposal for two main reasons. On the one hand, with the aim of repecting the structure of the data, maintaining order between observations is key to preserve efficiency and preventing abrupt transformations that may be unacceptable in practice in real contexts. Indeed, in the one-dimensional setting $X\in \mathbb{R}$, the continuous OT map between two probability measures is a non-decreasing function $T : \mathbb{R}^d \rightarrow \mathbb{R}^d$, taking into account the total ordering in the real line. For two or more dimensions, where there is not a total ordering, the monotonicity property (non-decreasing) is then generalized to that of cyclical monotonicity, which will be properly defined later in Section \ref{sec:theo_extended}. Therefore, a natural extension in higher dimensions $X\in \mathcal{X} \subset \mathbb{R}^d, \ d>1,$ should preserve such a property which is well suited to the fair learning problem. On the other hand, the fundamental difference with the other aforementioned approach lies in the simplicity of the assumptions required, since moment assumptions are not necessary, hence making it even more appropriate for application.



\subsection{Contributions}
\label{sec:contribution}
This paper introduces the \textit{Extended Total Repair (ExTR)},
an extension of the preprocessing technique of \cite{pmlr-v97-gordaliza19a} aimed at achieving fairness through the prism of OT theory.
Based on the smooth interpolation presented by \cite{10.1214/20-AOS1996}, we derive a consistent framework for obtaining unbiased attributes for new individuals.
Their solution to the cyclically monotone interpolation problem is built in a two-step procedure. First, they extend $T$ to a piecewise constant (hence not smooth) cyclically monotone map defined on a set in $\mathbb{R}^d$ whose complementary has Lebesgue measure zero. Second, they apply a regularization procedure yielding the required smoothness while keeping the interpolation feature.
To achieve this in an efficient way, we propose in this paper a combinatorial optimization method to compute specific parameters essential for defining the continuous OT map.
\par
Our approach can be delineated into two sequential phases that consists of formulating the dual problem and computing the minimum mean cycle of the associated graph through a scheme motivated by the use of the auction algorithm to solve the assignment problem. This procedure proves to be more efficient compared to the exact Hungarian algorithm. It is essential to note that the application framework is different, focusing specifically on the minimum mean cycle problem. Moreover, the continuous OT map is defined as the subdifferential of a convex function. In the process, we noticed that it is not convenient to compute the convex function and then its diferential. Instead of that, we suggest the stochastic subgradient descent method for computing the regularization of the interpolation function efficiently. Further details on all of the above sub-problems can be found in the Appendix \ref{app:sec:background}. 
\par
The interest of the proposed implementation method, which is an important contribution of this work, is twofold.
On the one hand, the procedure proposed can be used to integrate the repair into a K-fold CV procedure for the performance evaluation of an algorithm. Specifically, in each fold, the interpolation is computed for OT map between the groups in the training set, and then used to repair the remaining test set.
On the other hand, it can be employed when processing new (online) data once the model is deployed, assuming that distribution shift does not exist.
Alongside this paper, we present the code of all the computations performed with Python 3.8, available in our \href{https://github.com/emartindedi}{GitHub repository}. Besides, we also give different numerical experiments with simulated biased data and a real application whose objective is to forecast the customer's level of risk when granting a credit.

\subsection{Paper outline}
\label{sec:paper_outline}
After introducing the problem and reviewing the state of the art in Section \ref{sec:introduction}, the rest of the paper is organized as follows. Section \ref{sec:methodology} presents, on the one hand, two fundamental issues that are preliminary to the proposal of this work and, on the other hand, the concrete mathematical problem statement, which consists on defining a coherent function to approximate the repair values of some characteristics for new samples. Then the details to cover the implementation are explained in Section \ref{sec:computational_aspects}, where
we also provide insights into the adaptation of the method to real scenarios (Section \ref{sec:discussion_implementation}).
Afterward, in Sections \ref{sec:numericalexperiments} and \ref{sec:applications} we showcase the performance through numerical experiments and real data application with a benchmark dataset, respectively. 
Finally, Section \ref{sec:conclusions} concludes the paper and proposes some research lines for future investigation.
\par
As this paper requires knowledge of many subfields, we recommend that after Section \ref{sec:introduction}, a non-specialized reader first turn to the Appendix \ref{app:sec:background} and then continue with Section \ref{sec:methodology}. In particular, careful reading of Section \ref{app:sec:background:OT} is useful for understanding Section \ref{sec:methodology}.
Moreover, Sections \ref{app:sec:background:mcm} and \ref{app:sec:background:AP} contain background knowledge for our computational contribution provided in Section \ref{sec:computational_aspects}. Finally, in Appendix \ref{app-add-plots} additional details on the experiments can be found.

\section{Preliminaries and Problem Statement}
\label{sec:methodology}
This section is devoted, on the one hand, to introduce the two main preliminary aspects to our proposal of generalisation of the repair method for fair classification of new input data; and, on the other hand, to state its mathematical formulation. Therefore, in order to settle the notation and introduce some necessary concepts in this work, we will first briefly review in Section \ref{sec:total-repair} the total repair methodology of \cite{pmlr-v97-gordaliza19a} to obtain a fair representation of the data by means of the Wasserstein barycentre; and later in Section \ref{sec:prelim:cycmon} the smooth interpolation of the discrete OT map under cyclical monotonicity constraints of \cite{10.1214/20-AOS1996}. Then, in this context, we will explain in Section \ref{sec:theo_extended} how to gather both tools in order to construct the extended the total repair.


\subsection{Implementation of total repair}
\label{sec:total-repair}
\cite{pmlr-v97-gordaliza19a} established a framework to address fairness through the prism of OT theory for binary classifiers belonging to a family $\mathcal{G} = \{ g: \mathcal{X} \rightarrow \{0,1\} \}$. In this section we summarise the theoretical approach and its computational aspects that are necessary to understand our proposal. This approach involves modifying the input variables $X$ to create a modified version $\tilde{X} = T(X)$ in which the sensitive information has been deleted, so that any classifier $g \in \mathcal{G}$ trained on $(\tilde{X}, S)$ is fair with respect to the protected attribute $S$, that is, $DI(g, \tilde{X}, S) = 1$. 
The idea is to acquire a distribution $\mathcal{L}(\tilde{X})$ that closely resembles the distributions of the two subpopulations $\mu_{s}$, $s = 0, 1$, while $\tilde{X}\independent S$ and so $g(\tilde{X})\independent S$ for any classifier. The distribution proposed is the weighted Wasserstein barycenter between $\mu_{0}$ and $\mu_{1}$, which can be computed as: 
\begin{equation*}
\label{eq:5}
	\mu_{B} = ((1 - \lambda) Id + \lambda T)\# \mu_0,
\end{equation*}
where $\lambda = \frac{1}{2}$, $T: \mathbb{R}^{d} \rightarrow \mathbb{R}^{d}$ represents the OT map between $\mu_{0}$ and $\mu_{1}$ and $(G\# \mu )(A)=\mu (T^{-1}(A))$.
In particular, $\mu_{1}(B) = \mu_{0}(T^{-1}(B))$ for any measurable set $B \subset \mathbb{R}^{d}$.
The total repair approach is justified by means of an upper bound on the price of fairness for the transportation to the barycenter $\mu_{B}$  \citep[Theorem 3.3]{pmlr-v97-gordaliza19a}. 
To be more specific, it is demonstrated that the {\it minimal excess risk} $\mathcal{E}(T) = \inf_{g} R_{g}(\tilde{X}) - \inf_{g} R_{g}(X, S)$
, when considering the transformed data and the original data with the best classifier in each case (Bayes rules), is limited by the weighted Wasserstein variation of the conditional distributions multiplied by a constant. Here the risk is defined as $\mathcal{R}_{g}(Z)=P(g(Z)\neq Y)$, e.g. $Z=(X, S)$ or $Z=\tilde{X}$  that is the probability of a wrong classification.
\par
Hence, $\mathcal{L}(\tilde{X}) = \mu_{B}$, which means that the modified version of the original features are distributed according to the generalization for metric spaces of the mean of the conditional distributions.
For each $s \in \{0, 1\}$, if $\mu_{s}$ is absolutely continuous with respect to the Lebesgue measure in $\mathbb{R}^{d}$ and both $\mu_{s}$ and $\mu_{B}$ have finite second order moments, then there exists  the so called Brenier map $T_{s}: \mathcal{X}_{s} \longrightarrow \mathbb{R}^{d}$ \citep[Theorem 2.12]{villani2003topics}, which are  the unique OT maps $T_{s}$ between $\mu_{s}$ and $\mu_{B}$ with respect to the squared euclidean cost, for $s \in \{0, 1\}$. These maps are defined as the solution of the Monge problem, where $T_{s} \# \mu_{s} = \mu_{B}$. For  more details see  Appendix \ref{app:sec:background:OT}, in particular \eqref{eq:3}.
\par
Let $\{(\boldsymbol{x}_1,s_1,y_1), \ldots, (\boldsymbol{x}_n,s_n,y_n)\}$ be a sample i.i.d. from $(X,S,Y)$ and denote by $\mathcal{X}_{0} = \{ \boldsymbol{x}_{1}^{0}, ..., \boldsymbol{x}_{n_{0}}^{0} \}$ and $\mathcal{X}_{1} = \{ \boldsymbol{x}_{1}^{1}, ..., \boldsymbol{x}_{n_{1}}^{1} \}$ the observed characteristics of the two subgroups of the population, the privileged one ($S = 1$) and the unprivileged one ($S=0$), respectively, such that $n_0+n_1=n$.
In a practical setting, we have only access to the samples $\mathcal{X}_{s}$, not the entire distribution $\mu_{s}$ of each group $s = 0,1$. Hence, the quadratic cost function takes a discrete form, expressed as a matrix $C = (c_{i,j})$, where $c_{i,j} = \Vert \boldsymbol{x}_{i}^{0} - \boldsymbol{x}_{j}^{1} \Vert^{2}$, $1 \leq i \leq n_{0}$, $1 \leq j \leq n_{1}$. 
Considering the empirical measures: 
\begin{equation*}
	\mu_{0, n} = \sum_{i=1}^{n_{0}}\frac{1}{n_{0}}\delta_{\boldsymbol{x}_{i}^{0}} \quad \text{and} \quad \mu_{1, n} = \sum_{j=1}^{n_{1}}\frac{1}{n_{1}}\delta_{\boldsymbol{x}_{j}^{1}},
\end{equation*}
where $\delta_{\boldsymbol{x}}$ is the Dirac measure, the Wasserstein distance 
between $\mu_{0, n_0}$ and $\mu_{1, n_1}$ is the squared root of the optimum of a network flow problem known as the transportation problem \citep{pmlr-v32-cuturi14}. It is well known that the OT solution is achieved computing  a matrix $\gamma$ that minimizes the transportation cost between the two empirical distributions, that is,
\begin{equation*}
\begin{cases}
   \min_{\gamma} \sum_{\substack{1 \leq i \leq n_{0},\\1 \leq j \leq n_{1}}} c_{i,j}\gamma_{i,j},  \text{    subject to:}\\
  \gamma_{i,j} \geq 0,\\
  \sum_{i=1}^{n_{0}} \gamma_{i,j} = \frac{1}{n_{1}},  \quad \forall j \in \{1, ..., n_{1}\},\\
  \sum_{j=1}^{n_{1}} \gamma_{i,j}= \frac{1}{n_{0}}, \quad \forall i  \in \{1, ..., n_{0}\} .\\
\end{cases}       
\end{equation*}
If $\Hat{\gamma}$ is a solution to the above linear program, then the measure:
\begin{equation*}
	\mu_{B, n} = \sum_{\substack{1 \leq i \leq n_{0},\\1 \leq j \leq n_{1}}} \Hat{\gamma}_{i,j} \delta_{\pi_{0}\boldsymbol{x}_{i}^{0} + \pi_{1}\boldsymbol{x}_{j}^{1}}
\end{equation*}
is an empirical barycenter of $\mu_{0, n}$ and $\mu_{1,n}$, with weights $\pi_{0}$ and $\pi_{1}$.
To sum up, the computation of the repaired data set relies on the transportation matrix $\Hat{\gamma}$ from $\mathcal{X}_{0}$ to $\mathcal{X}_{1}$.

\subsection{Smooth interpolation of the OT map under cyclical monotonicity constraints}
\label{sec:prelim:cycmon}

The challenge lies in creating an approximation that effectively generalizes the transformation to new data while maintaining statistical consistency. This leads us to question which characteristics should be preserved by the extension of the discrete OT map. 
In the one dimensional setting, the OT map between two probability measures is a non-decreasing function $T$ such that $T\#\mu_{0} = \mu_{1}$.
For more than one dimension, cyclical monotonicity  is a generalization of the notion of monotonicity to vector-valued functions, which is defined as follows:


\begin{definition}
\label{def:1}
Let $\mathcal{X}, \mathcal{\tilde{X}} \subset \mathbb{R}^d$. A subset $\Upsilon$ of $\mathcal{X} \times \mathcal{\tilde{X}}$ is said to be {\it cyclically monotone set} if, for any finite collection of points $\ \{ (\bm{x_{i}}, \bm{\tilde{x}_{i}}) \}_{i = 1}^{k} \subseteq \Upsilon$, 
\begin{equation*}
    \langle \bm{\tilde{x}_{1}} , \; \bm{x_{2}} - \bm{x_{1}} \rangle  + \langle \bm{\tilde{x}_{2}} , \; \bm{\tilde{x}_{3}} - \bm{x_{2}} \rangle + ... + \langle \bm{\tilde{x}_{k}} , \; \bm{x_{1}} - \bm{x_{k}} \rangle \leq 0,
\end{equation*}
where $\langle \cdot , \; \cdot \rangle $ stands for an inner product and $k = 2, 3, ..., |\Upsilon|$.

A mapping $F: \mathbb{R}^{d} \rightarrow \mathbb{R}^{d}$ is said to be \textit{cyclically monotone mapping} if and only if its graph $\{(\boldsymbol{x}, F (\boldsymbol{x})) \text{  }|\text{  } \boldsymbol{x} \in \mathbb{R}^{d}\}$ is cyclically monotone. 
\end{definition}

The cyclical monotonicity is essentially a set property, so that if $\Upsilon$ is cyclically monotone, then any subset of $\Upsilon$ is also cyclically monotone.
Notice that when $\phi: X \rightarrow [-\infty,+\infty]$ is a convex function, then the subgradient, given by $\partial\phi = \{(x, \tilde{x}): \phi(z) - \phi(x) \geq \langle \tilde{x}, z - x\rangle, \forall z\} \subset X \times Y$, is cyclically monotone. The converse is shown by \cite[Theorem 1]{Rockafellar1966CharacterizationOT} :
Let $\Upsilon \subset X \times Y$ be a cyclically monotone set, then there exists a convex function $\phi: X \rightarrow \mathbb{R}$ such that $\Upsilon \subset \partial\phi$.
We refer to \cite{kausamo202360}, for a survey of the notion of cyclic monotonicity.

\par
Hence, a natural extension of the discrete OT map in higher dimensions should preserve the cyclical monotonicity property; or equivalently, its graph should exhibit cyclical monotonicity.
Therefore, in our particular problem, the maps $T_{s}$ satisfy the cyclical monotonicity property for each $s = 0, 1$, as they are the gradient of a convex function ($\{(\boldsymbol{x}, T_{s} (\boldsymbol{x})) \text{  }|\text{  } \boldsymbol{x} \in \mathcal{X}_{s}\}$ is cyclically monotone).
%
To broaden the univariate concept of a distribution function to multivariate contexts, \cite{10.1214/20-AOS1996} introduced a novel approach inspired by McCann's main theorem \citep[p. 310]{10.1215/S0012-7094-95-08013-2}. The authors provided the first continuous extension consisting of a smooth interpolation function maintaining the cyclical monotonocity.
Later, \cite{delara2021consistent} demonstrated that the proposed interpolation is consistent in the sense that the resulting transport plans asymptotically converge to the continuous OT map as the sample size increases.
We subsequently employ this smooth interpolation under cyclical monotonicity constraints to establish a consistent framework for extending the total repair bias mitigation technique of \cite{pmlr-v97-gordaliza19a}.

\subsection{Theoretical aspects of the Extended Total Repair}
\label{sec:theo_extended}
\par
Once the total repair process is completed, a sample $\tilde{\mathcal{X}}_{0}\cup\tilde{\mathcal{X}}_{1} $ from the Wasserstein barycenter $\mu_{B} = T_{s}\# \mu_{s}$ is obtained, where $\tilde{\mathcal{X}}_{s}$ are the repaired subsamples of each group $s = 0,1$.
Our primary goal is to extend the discrete map $T_{s}: \mathcal{X}_{s} \rightarrow \tilde{\mathcal{X}_{s}}$ to a continuous function $\Bar{T}_{s}: \mathbb{R}^{d} \rightarrow \mathbb{R}^{d}$ outside the support of the empirical distribution. 
This extension enables us to efficiently generate the repaired version of a new set of points, without computing the transport plan again. Instead, we propose to simply apply the function $\Bar{T}_{s}$ based on the class $s$ to which the sample belongs.
\par
For ease of notation, given that the construction of the interpolation function is group-agnostic, we fix the class $s$ and no longer write the superscript indicating the points' subgroup in $\mathcal{X}_{s}$ or $\tilde{\mathcal{X}}_{s}$.
The goal is to construct a smooth (at least continuous) cyclically monotone map $\Bar{T}_{s}: \mathbb{R}^{d} \rightarrow \mathbb{R}^{d}$ such that $\Bar{T}_{s}(\boldsymbol{x}_{i}) = T_{s}(\boldsymbol{x}_{i}) = \tilde{\boldsymbol{x}}_{i}, \ i = 1, ..., n_{s}$.
It is well-known that the subdifferential of a convex function from $\mathbb{R}^{d}$ to $\mathbb{R}$ exhibits cyclical monotonicity.
As a generalization of the classic result by \cite{Rockafellar1966CharacterizationOT} was made by \cite{10.1214/20-AOS1996} establishing the converse: any cyclically monotone subset $S = \{ (\boldsymbol{x}_{i}, \tilde{\boldsymbol{x}}_{i}) \text{  }|\text{  } i = 1, ..., n \}$ of $\mathbb{R}^{d} \times \mathbb{R}^{d}$ is contained in the subdifferential of some convex function.
The construction of the function primarily focuses on convex functions that are differentiable and adopts a two-step approach: 
\begin{itemize}
    \item[]\textbf{(Step 1)} The map $T_s$ will be extended to a piecewise constant cyclically monotone map defined on a set in $\mathbb{R}^{d}$ whose complementary has Lebesgue measure zero. 
    In detail, due to the cyclically monotone property \citep[Proposition 3.1]{10.1214/20-AOS1996}, there exist real numbers $(\psi_{1}, ... ,\psi_{n})$ such that:
	\begin{equation}
    	\label{eq:6}
    	\langle \boldsymbol{x}, \tilde{\boldsymbol{x}}_{i} \rangle - \psi_{i} > \max_{j \neq i} (\langle \boldsymbol{x}, \tilde{\boldsymbol{x}}_{j} \rangle - \psi_{j}) \; \; \; \text{  } i = 1, ..., n.
	\end{equation}
	Hence,  the convex function
	\begin{equation}
	\label{eq:7}
    	\tilde{\varphi}_{n}(\boldsymbol{x}) =  \max_{1\leq j \leq n} \{ \langle \boldsymbol{x}, \tilde{\boldsymbol{x}}_{j} \rangle - \psi_{j}  \},
	\end{equation}
	 can be extended to $\Bar{T}_{s}(\boldsymbol{x}) = \nabla \tilde{\varphi}_{n}(\boldsymbol{x})$, $\boldsymbol{x} \in \mathbb{R}^{d}$. 
	It should be noted that when $\boldsymbol{x} \in \bigcup_{i=1}^{n}C_{i}$, being $C_{i} = \{\boldsymbol{x} \in \mathbb{R}^{d} | (\langle 	\boldsymbol{x}, \tilde{\boldsymbol{x}}_{i} \rangle - \psi_{i}) > \max_{j \neq i} (\langle \boldsymbol{x}, \tilde{\boldsymbol{x}}_{j} \rangle - \psi_{j})\}$ convex sets, then the function $ 	\tilde{\varphi}_{n}$ is differentiable, furthermore, $\nabla \tilde{\varphi}_{n}(\boldsymbol{x}) = \tilde{\boldsymbol{x}}_{i}$, $\boldsymbol{x} \in C_{i}$ and the complement of $\bigcup_{i=1}^{n}C_{i}$ has Lebesgue measure zero, concluding the first step.

    \item[] \textbf{(Step 2)} As this map is piecewise constant, it cannot be considered as smooth. To overcome this limitation, a regularization process is implemented in step two that confers the required smoothness, while preserving the interpolation property. To obtain a regular interpolation, preserving the cyclical monotonicity, defined in the space $\mathbb{R}^{d}$, the Moreau-Yoshida regularization of $\tilde{\varphi}_{n}$ \citep{rockafellar2009variational} will be considered, namely:
    \begin{equation}
        \label{eq:MYreg}
            \varphi_{\epsilon}(\boldsymbol{x}) = \inf_{\tilde{x} \in \mathbb{R}^{d}}\left\{ \tilde{\varphi}_{n}(\boldsymbol{\tilde{x}}) + \frac{1}{2 \epsilon} \| \boldsymbol{\tilde{x}} - \boldsymbol{x} \|_2^2 \right\}, \text{         } \boldsymbol{x} \in \mathbb{R}^{d}, \epsilon > 0.
    \end{equation}
 Then, the function $\Bar{T}_{s}(\boldsymbol{x}) = \nabla \varphi_{\epsilon}(\boldsymbol{x}):\mathbb{R}^{d} \mapsto\mathbb{R}^{d}$ provides a cyclically monotone interpolation that is continuous and Lipschitz with constant $\frac{1}{\epsilon_{0}}$, where
	\begin{equation}
	\label{eq:8}
	\epsilon_{0} = \frac{1}{2} \min_{1 \leq i \leq n_{s}} ((\langle \boldsymbol{x}_{i}, \tilde{\boldsymbol{x}}_{i} \rangle - \psi_{i}) - \max_{j \neq i} (\langle \boldsymbol{x}_{i}, \tilde{\boldsymbol{x}}_{j} \rangle - \psi_{j})) \ \text{\citep{10.1214/20-AOS1996}} .
	\end{equation}
	Finally, the smoothest possible interpolation is achieved by taking the weights $\psi_{i}$ that satisfy the condition \eqref{eq:6} and optimize $\epsilon_{0}$ in Eq. \eqref{eq:8}. 
	As a consequence, the optimal smoothing value is half of the maximum in the linear program: 
	\begin{equation}
	\label{eq:9}
    		\max_{\psi, \epsilon} \, \epsilon \, \, \, \, \, \,\text{s.t.} \, \, \,\, \, \, \langle \boldsymbol{x}_{i}, \tilde{\boldsymbol{x}}_{i} - \tilde{\boldsymbol{x}}_{j} \rangle \geq \psi_{i} - \psi_{j} + \epsilon,  \, \, \,i, j \in \{1, ..., n\}, i \neq j.
	\end{equation}
	This optimization problem will be addressed comprehensively in the Section \ref{sec:combinatorial_optimization}. 
\end{itemize}

To sum up, for such optimal smoothing values, $ \Bar{T}_{s}(\boldsymbol{x}) = \nabla\varphi_{\epsilon_{0}}(\boldsymbol{x}), \forall \boldsymbol{x} \in \mathbb{R}^{d}$, is the interpolation function of interest from $\mathbb{R}^{d}$ to $\mathbb{R}^{d}$.
To get a clearer picture of the usefulness of the interpolation in a K-fold CV procedure, it may be useful to have a look at some of the results on a simulated (one-dimensional) biased dataset that will be properly introduced and studied later in the paper (Experiment $\mathcal{E}$1 - A of Section \ref{sec:numericalexperiments}). Precisely, in each of the $K$ splits, we can use the training set to calculate the repair transport plan and interpolate it to be able to repair the test set. 
First of all, the training set is divided into the two subsamples by group, namely $\mathcal{X}_s^{train},$
from which the transportation plans $T_s$ and their interpolations $\Bar{T}_{s}$ are obtained, so that the interpolated points for the test set $\mathcal{X}_s^{test}$ can be computed. The graphs of such functions are represented in Figure \ref{im:1} : $T_0$ in blue, $T_1$ in green, $\Bar{T}_{0}$ in pink, and $\Bar{T}_{1}$ in orange color.
In the following section, we explain the computational aspects of our proposal to efficiently calculate $\Bar{T}_{s}$.



\section{Computational aspects of the Extended Total Repair}
\label{sec:computational_aspects}
\par
Significant computational challenges arise when it comes to the calculation of the smoothest interpolation $\Bar{T}_{s}$, due to the complexity of the optimization problems involved. 
Our proposal is structured in two different steps. Firstly, in Section \ref{sec:combinatorial_optimization} the dual of problem \eqref{eq:9} is reduced to the computation of a minimum cycle mean (m.c.m.) of the associated graph. Secondly, from this solution, the Moreau-Yosida regularization in \eqref{eq:MYreg} is optimized through stochastic subgradient descent in Section \ref{sec:Moreau-Yoshida}.

\subsection{A combinatorial optimization approach for computing the optimal smoothing values}
\label{sec:combinatorial_optimization}
\par
We propose a computationally efficient method for the calculus of the optimal smoothing value $\epsilon_{0}$, which is exactly $\frac{1}{2}$ times the solution of Eq. \eqref{eq:9}.
The approach can be delineated into two sequential phases: (i) Formulate the dual problem; and (ii) Compute the m.c.m. of the associated graph. As will be explained later, the idea behind the second phase is motivated by the use of the auction algorithm to solve the {\it assignment problem} (AP). This adaptation is an important contribution of this paper, which proves to be more efficient compared to the exact Hungarian algorithm. It is essential to note that the application framework is different, focusing specifically on the m.c.m. problem. We refer to Sections \ref{app:sec:background:mcm} and \ref{app:sec:background:AP} in the Appendix, for prior knowledge on these topics.

\par
Let $G = (V, E)$ represent a directed graph with $n$ nodes and $m$ arcs. The graph's edges have associated weights given by the function $f: E \longrightarrow \mathbb{R}$. For simplicity, we express these weights as a square matrix $C \in \mathcal{M}_{n}(\mathbb{R})$ where the entries are defined as $ c_{i,j} = f(e_{i,j}) $ with $e_{i,j}$ representing the edge between vertex $i$ and $j$. 
In our context, we consider $c_{i,j} = \langle \boldsymbol{x}_{i}, \tilde{\boldsymbol{x}}_{i} - \tilde{\boldsymbol{x}}_{j} \rangle$ for $i, j \in \{1, ..., n\}$.
To further clarify, we introduce the constant $C_{bound} = 1 + \max{\{|c_{i,j}| : (i,j) \in E \} }$.
Notice that the matrix $C = (c_{i,j})_{i,j}^{n}$ possesses a diagonal comprised entirely of zeros by explicit definition. This fact implies, within the framework of the AP, that individual $i$ cannot be assigned the object $i$. 
%
Moreover, we shall make the following three assumptions: 
\begin{itemize}
\item[($\mathcal{A}$1)]  The cost matrix in the AP is squared. If not, it is converted by adding zeros.
\item[($\mathcal{A}$2)] The graph $G$ is strongly connected (i.e. every vertex is reachable from every other vertex). Otherwise, the m.c.m. would be the minimun m.c.m. among the strong components of $G$, which can be found in $\mathcal{O}(n + m)$.
\item[($\mathcal{A}$3)] The AP has a feasible solution. The infeasibility of the AP can be detected in $\mathcal{O}(\sqrt{n} m)$ time by the bipartite cardinality matching algorithm of \cite{doi:10.1137/0202019}.
\end{itemize}
%
\subsubsection{Dual problem formulation}
The dual formulation of the optimization problem \eqref{eq:9} is a circulation problem over a complete graph $G$ with $n$ vertices, that is:
\begin{equation}
\label{eq:10}
\begin{array}{ll@{}ll}
& \min_{z_{i,j}, i \neq j} \displaystyle\sum\limits_{i,j=1,...,n; i \neq j} z_{i,j}\langle \boldsymbol{x}_{i}, \tilde{\boldsymbol{x}}_{i} - \tilde{\boldsymbol{x}}_{j} \rangle & &\\
\text{subject to}& \displaystyle\sum\limits_{j=1,...,n; i \neq j} (z_{i,j} - z_{j,i}) = 0,  \quad &i, j=1 ,\dots, n\\
                 & \displaystyle\sum\limits_{i,j=1,...,n; i \neq j} z_{i,j} = 1, \quad &i, j=1 ,\dots, n\\
                 &  z_{i,j} \geq 0,  &i, j=1 ,\dots, n.\\
\end{array}
\end{equation}

\par
By the Flow Decomposition Theorem \citep[Theorem 3.5]{Ahuja_1993_9423}, any circulation is of the form $z_{i,j} = \sum_{W \in \mathcal{W}}\delta_{i,j}(W)f(W),$
where $\mathcal{W}$ denotes the set of all cycles in the graph and $f(W) \geq 0$ is the flow along cycle $W$.  
Also, 
\begin{equation*}
\begin{cases}
  \delta_{i,j}(W) = 1  \quad \text{if the arc connecting } i \text{ and } j \text{ belongs to cycle } W,\\
  \delta_{i,j}(W) = 0 \quad \text{otherwise. }\\
\end{cases}       
\end{equation*}
Writing $c_{i,j} = \langle \boldsymbol{x}_{i}, \tilde{\boldsymbol{x}}_{i} - \tilde{\boldsymbol{x}}_{j} \rangle$ and $c(W)= \sum_{i,j}\delta_{i,j}(W)c_{i,j}$, the objective function can be expressed as:
\begin{equation*}
    \sum_{i,j=1,...,n; i \neq j} c_{i,j} z_{i,j} = \sum_{W \in \mathcal{W}} c(W)f(W),
\end{equation*}
subject to $\sum_{W \in \mathcal{W}} |W| f(W) = 1$, where $|W|$ denotes the length of the cycle $W$. 
Being $\tilde{f}(W) = |W| f(W)$, the program can be rewritten as:
\begin{equation*}
    \min_{\tilde{f}(W)} \sum_{W \in \mathcal{W}} \tilde{f}(W) \frac{c(W)}{|W|} \, \, \, \, \, \,\text{s.t.} \, \, \,\, \, \, \sum_{W \in \mathcal{W}} \tilde{f}(W) = 1, \,\, \tilde{f}(W) \geq 0.
\end{equation*}
Consequently, the optimal solution to \eqref{eq:10} is $z_{i,j} = \delta_{i,j}(\Hat{W})/|\Hat{W}|$, where $\Hat{W}$ is a minimum mean cost cycle, or in other words, the cycle with the minimum cost when compared to all other cycles of $\frac{c(W)}{|W|}$. The next phase focuses on calculating such a minimum, which we will denote from now on by $\epsilon^{\star}$. 

\subsubsection{Minimum cycle mean computation through the application of the auction algorithm to the assignment problem}
The problem of finding the m.c.m. has been a significant challenge in the field of combinatorial optimization and, therefore, the subject of extensive algorithmic research for many years.
A feasible approach is the characterization by Karp's algorithm \citep{karp1978characterization} yet with running time $\mathcal{O}(n^{3})$, being $n$ the number of nodes of the corresponding graph.
%
%
The most advanced algorithm presently available for this problem was developed by \cite{Orlin1992} taking $\mathcal{O}(\sqrt{n} |E| log(n C_{bound}))$ time. 
The method consists in an approximate binary search, as in \cite{Zemel1987}.
In particular, once defined the search interval $[L, U]$ for the optimum $\epsilon^{\star}$, it proposes an application of the cost scaling phase in an hybrid assignment algorithm, instead of executing a sequence of shortest path problems.
As a result, the scaling method for the m.c.m. entails moving the framework from a fundamental graph algorithmic problem of computing the
m.c.m. on a finite directed graph, to that of a fundamental combinatorial optimization problem such as the weighted matching in a bipartite graph, also referred to as the assignment problem (AP). Recall that (see Appendix \ref{app:sec:background:AP}) this problem consists of finding a matching of a given size in a weighted graph, in which the sum of weights of the edges is minimum. 
%
\par
On the basis of a high level description of the algorithm provided by \cite{Orlin1992}, we derive a consistent and efficient framework for efficiently computing the m.c.m. of a graph through a strategy for solving the AP. In particular, when this problem is posed as a linear program, a solution can be obtained in polynomial time \citep{zbMATH05522909}.
However, the classical approach for addressing this problem does not readily lend itself to parallelization. 
To address this limitation, rooted in the economic concept of competitive equilibrium, \cite{Bertsekas_auction} alternatively proposes to apply the auction algorithm running in pseudo-polynomial time.
\par
Inspired by this idea, our proposal combines the auction algorithm with the successive shortest path procedure, enhancing the computational efficiency beyond what can be achieved by using either technique alone.
Moreover, the practical performance of the algorithm proposed is improved due to the idea of scaling. It consists of applying the algorithms several times, starting with a large value of the objective function and successively reducing until it is less than some critical value (in our case $1/n^{2}$). Each application of the algorithm provides good initial values for the next application, while maintaining the optimality conditions.

\par
\begin{figure}[h!]
	\centering
    \includegraphics[width=.5\textwidth]{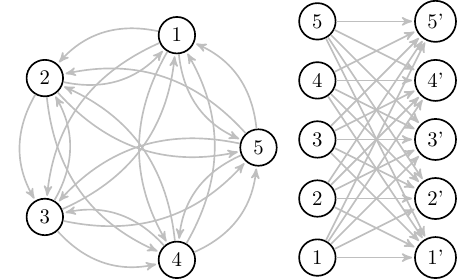}
	\caption{Scheme of the node splitting technique for the value of $n = 5$.}
	\label{fig:node-splitting}
\end{figure}

The first step in implementing our proposal is to establish the equivalence between the m.c.m. and the AP problem, for which we will use the node-splitting technique represented in Figure \ref{fig:node-splitting}. This involves splitting every node $i$ within the set $N_{1} = \{1, 2, ..., n\}$ into two different nodes denoted as $i$ and $i^{\prime}$. Furthermore, each edge $e_{i,j}$ in the original graph is substituted with the edge $e_{i, j^{\prime}}$ and the costs associated remains the same.  
It is necessary to incorporate supplementary edges $e_{i, i^{\prime}}$ in the transformed graph, with costs $\delta$. 
As a result of these transformations, an AP emerges characterized by the sets $N_{1}$ and $N_{2} = \{1^{\prime}, 2^{\prime}, ..., n^{\prime}\}$.
\par
In addition, two further remarks are of interest before presenting the algorithm. On the one hand, the formulation of the m.c.m. problem does not impose a requirement that the directed cycles under consideration must all have the maximum possible length $n$, whose equivalent concept will be a $0-1$ assignment $\boldsymbol{x}$. This observation underscores the utilization of partial assignments, where certain nodes within the problem may remain unassigned. Formally, a partial assignment $\boldsymbol{x}$ is a solution of the assignment problem for which $\sum_{\{j: (i,j) \in A\}} x_{i,j} \leq 1$ for some set of pairs of subscripts $A$. On the other hand, in a generic scenario, the entries of the matrix known do not need to be greater or equal than zero. Nevertheless, for the algorithm proposed it is a necessary condition. This is solved by adding a suitable large constant value to all the entries (arc costs) without altering the solution. Moreover, it will be useful that the entries are also integer numbers instead of float.

\par
We finally present in Algorithm  \ref{alg:mmc} our proposal for obtaining the m.c.m. efficiently. 
For a comprehensive understanding of the selected parameters, including proofs and detailed explanations, we refer to \cite{Orlin1992}. At a higher level of abstraction, the algorithm
initiates by establishing an interval within which we ascertain the presence of $\epsilon^{\star} \in [ LB, UB ]$, where $ LB = - C_{bound}$ and $UB = C_{bound}$. 
Subsequently, the methodology employed diverges from conventional binary search procedures. In this case, each iteration entails the resolution of an assignment problem. 
In order to obtain an optimal solution, an hybrid version of the auction algorithm and the successive shortest path algorithm is applied until the difference $UB-LB$ becomes small, namely, $|UB-LB| < \frac{1}{n^{2}}$. 
Finally, through the assignment $x$ founded, we can return to the original problem and define the cycle with minimum mean cost among all directed cycles in the graph with arc costs defined through the matrix $C = (c_{i,j})_{i, j = 1, ..., n} = (\langle \boldsymbol{x}_{i}, \tilde{\boldsymbol{x}}_{i} - \tilde{\boldsymbol{x}}_{j} \rangle)_{i, j = 1, ..., n}$. 

\begin{algorithm}[h!]
\SetAlgoNoLine
\caption{Minimum mean cycle computation}\label{alg:mmc}
\KwInput{Matrix $C \in \mathcal{M}_{N}$
}

\KwOutput{$\epsilon^{\star} = m(C^{\star}_{cycle}) \in [LB, UB]$ the minimum mean cost of the cycle founded.}

\textbf{Initialization step:}

\Indp

treat-the-matrix-if-needed(C)

define-parameters\;

\Indp 
$ C_{bound} = 1 + \max{\{|c_{i,j}| : (i,j) \in A \} } $\;
$ LB = - C_{bound}$ and $UB = C_{bound}$\;
node potentials: $\pi(i) = \frac{-C}{2}$ for all $i \in N_{1}$ and $\pi(j) = 0$ for all $j \in N_{2}$\;
\Indm

let $x$ be the null assignment\;

\Indm

\While{$ UB-LB \geq \frac{1}{n^{2}}$}{

$\delta = \frac{1}{2}(UB + LB)$\;
$\epsilon = \frac{1}{8}(UB - LB)$\;
$k = 3$\;
$L = 2 (k + 1)[\sqrt{n}]$\;
$C_{\delta} $ : matrix $C$ but with $\delta$ as the values in the principal diagonal\;

\textbf{auction-algorithm} ($C_{\delta}, k, \epsilon, L, \text{node potentials}, x$) \;

\Indp 

$\pi(j) = \pi(j) + k \cdot \epsilon$ for all $j \in N_{2}$\;
\While{there is an eligible unassigned node $i \in N_{1}$}{

calculate the reduced cost matrix $\tilde{c}_{i,j} = c_{i,j} - \pi(i) + \pi(j)$\;

\uIf{there is an admissible arc ($i_{\text{unassigned}}, j$)}{

$\pi(j) = \pi(j) + \epsilon$ \;
\uIf{node $j$ was already assigned to some node $l \in N_{1}$}{
deassign node $l$ from node $j$\;
}
assign node $i_{\text{unassigned}}$ to node $j$\;

}

\uElse{
$\pi(i_{\text{unassigned}}) = \pi(i_{\text{unassigned}}) + \epsilon$ \;
}

}

\Indm

\textbf{successive-shortest-path} ($C_{\delta}, k, \epsilon, L, \text{node potentials}, x $)

\Indp 

\While{there are unassigned nodes in $N_{1}$}{

calculate the reduced cost matrix $\tilde{c}_{i,j} = c_{i,j} - \pi(i) + \pi(j)$\;
define $d_{i,j} = \max{\{ 0, \frac{\tilde{c}_{i,j}}{\epsilon} + 1 \}}$\;
select an unassigned node $i_{\text{unassigned}} \in N_{1}$\;
apply Dijkstra's algorithm with $d_{i,j}$ as arc lengths with source $i_{\text{unassigned}}$ until some unassigned node $j_{l}$ is permanently labeled\;
let $w[i]$ denote the distance labels of nodes that are permanently labeled\;
update $\pi_{i} = \pi_{i} + \epsilon (w[j_{l}] - w[i])$ for all permanently labeled nodes $i \in N_{1}$
augment one unit of flow along the shortest path from node $i_{\text{unassigned}}$ to $j_{l}$ and update the assignment $x$\;
}

update the assignment and the node potentials\;

\Indm

\uIf{$x$ is a uniform assignment}{
$LB = \delta - 2 \cdot \epsilon $\;
}
\uElse{
$UB = \delta + 2 \cdot \epsilon$\;
break\;
}

}

from the non uniform assignment $x$, construct the minimum mean directed cycle $C^{\star}_{cycle}$ \;

\end{algorithm}


\subsection{Optimizing Moreau-Yoshida regularization with stochastic subgradient descent}
\label{sec:Moreau-Yoshida}
\par
The explicit calculation of the function $\Bar{T}_{s}$ is limited by the inefficiency of directly calculating the function $\varphi_{\epsilon_{0}}$ and its corresponding gradient. The work on maximal monotone operators of \cite{yosida1967functional} contributes in this sense, allowing to characterize the function $\Bar{T}_{s} = \nabla\varphi_{\epsilon_{0}}$ in terms of the Moreau envelope \citep{rockafellar2009variational}. Precisely, the Moreau envelope of an extended real-valued function $g:\mathbb{R}^{d} \rightarrow \mathbb{R} \cup \{ \infty \}$, also called Moreau-Yoshida approximate, is defined as
\begin{equation*}
    \mathrm{env}_{\epsilon g}(\boldsymbol{x}) =
        \inf_{\boldsymbol{\tilde{x}} \in \mathbb{R}^{d}} \left\{ g(\boldsymbol{\tilde{x}}) + \frac{1}{2 \epsilon} \| \boldsymbol{x} - \boldsymbol{\tilde{x}} \|_2^2 \right\},
\end{equation*}
and has been extensively studied to explore its regularization properties, both theoretically and by using algorithmic results. 
Under general conditions, $\mathrm{env}_{\epsilon g}$ is a $\mathcal{C}^{1}$ with Lipschitz continuous gradient, and critical points of $g$ are fixed points of the proximal mapping: 
\begin{equation}
    \label{eq:11}
    \mbox{prox}_{\epsilon g}(\boldsymbol{x}) = \operatorname{argmin}_{\boldsymbol{\tilde{x}} \in \mathbb{R}^{d}} \left\{ g(\boldsymbol{\tilde{x}}) +
            \frac{1}{2\epsilon} \| \boldsymbol{\tilde{x}}-\boldsymbol{x} \|_2^2 \right\}.
\end{equation}
Finally, from Eq. \eqref{eq:MYreg}, and
according to \cite[Theorem 2.26]{rockafellar2009variational}, the following characterization of $\Bar{T}_{s}$  is obtained
\begin{equation}
    \label{eq:12}
    \Bar{T}_{s}(\boldsymbol{x}) = \nabla \varphi_{\epsilon_{0}}(\boldsymbol{x}) = \frac{1}{\epsilon_{0}} (\boldsymbol{x} - \mbox{prox}_{\epsilon_{0} \text{ }\tilde{\varphi}_{n}}(\boldsymbol{x}) ),
\end{equation}
where $\epsilon_{0}$ is half of the optimal value of \eqref{eq:9}.
\par
Due to its properties, the proximal point mapping is the basis of many optimization techniques for convex functions. In particular, when $g$ is convex lower semi-continuous and proper, the proximal mapping is a
maximal monotone operator and its fixed points are the minima of $g$ \citep{Lucet_Moreau}. 

To sum up, in order to complete our ExTR procedure, the unbiased version of a new observation $\boldsymbol{x} \not\in \mathcal{X}_{s}$ will be $\tilde{\boldsymbol{x}}=\Bar{T}_{s}(\boldsymbol{x})$, obtained according to Eq. \eqref{eq:12}. Therefore, the problem is reduced to a convex optimization problem \eqref{eq:11} with $g = \tilde{\varphi}_{n}$ and $\epsilon = \epsilon_{0}$, which can be efficiently solved through the Algorithm \ref{alg:stochastic} proposed.
\par
Let the objective function (note that $\boldsymbol{x}$ is fixed for the moment) be
\begin{equation}
    \label{eq:13}
    h(\boldsymbol{\tilde{x}}) = \tilde{\varphi}_{n}(\boldsymbol{\tilde{x}}) +
            \frac{1}{2\epsilon} \| \boldsymbol{\tilde{x}}-\boldsymbol{x} \|_2^2 \;\; \text{, } \boldsymbol{\tilde{x}} \in \mathbb{R}^{d}. 
\end{equation} 
Since $\tilde{\varphi}_{n}$ is convex, the function $h$ is $\frac{1}{\epsilon_{0}}$ - strongly convex. Hence, the infimum is actually a minimum and $\mbox{prox}_{\epsilon_{0} \tilde{\varphi}_{n}}(\boldsymbol{x})$ is unique for a given $\boldsymbol{x} \in \mathbb{R}^{d}$. 
In this case, $h$ is not necessary a differentiable function due to the fact that the only requirement for the map $\tilde{\varphi}_{n}$ is the convexity. Nevertheless, the map $\varphi_{\epsilon_{0}} = \mathrm{env}_{\epsilon_{0} \tilde{\varphi}_{n}}$ is continuous and differentiable. 
\par
As a consequence, the gradient descent algorithm \citep{Goodfellow-et-al-2016} cannot be applied directly, but the update step can be adapted making use of the subgradient instead, having equivalently convergence results.
Besides that, the direction of the update does not need to be the subgradient, it is sufficient that the expectation of the random vector selected belongs to the subgradient. 
All definitions and notions about convex analysis are taken from \cite[Chapter 12]{10.5555/2621980}.

\begin{algorithm}[H]
\caption{Stochastic Sub-gradient Descent to minimize $h(\boldsymbol{\tilde{x}})$}\label{alg:stochastic}
\KwInput{Learning rate $\eta_{0} = \epsilon_{0} > 0$, integer number $T >0$, tolerances $rtol1, rtol2 >0$.}
\KwOutput{$\boldsymbol{\tilde{x}} = \boldsymbol{\tilde{x}}^{(T)}$ }
$\boldsymbol{\tilde{x}}^{(0)} \gets \boldsymbol{x}_{1}^{s} \; \text{ (first element of } \; \mathcal{X}_{s} \text{, where } s \text{ is the class of the fixed new value } \boldsymbol{x} \text{)}$ \;
Initialize $t=1$\;
\While{(1) $h(\boldsymbol{\tilde{x}}^{(t)}) < rtol_{1}$, and (2) $h(\boldsymbol{\tilde{x}}^{(t)}) < h(\boldsymbol{\tilde{x}}^{(t+1)})$, and (3) $|\partial h(\boldsymbol{\tilde{x}}^{(t)})| < rtol_{2}$, and (4) $t \leq T$}{
  Choose $\boldsymbol{v}_{t}$ at random from $\partial h(\boldsymbol{\tilde{x}}^{(t)})$\;
  Update $\boldsymbol{\tilde{x}}^{(t+1)} = \boldsymbol{\tilde{x}}^{(t)} - \eta_{t} \boldsymbol{v}_{t}$\;
  Update learning rate $\eta_{t+1} = \frac{\epsilon_{0}}{t+1}$\;
  Update $t=t+1$ \;
}
\end{algorithm}

Thus, the computation of the gradient of $h$ at each iteration $t$
is computed as
\begin{equation*}
	\partial h(\boldsymbol{\tilde{x}}^{(t)}) = \partial \tilde{\varphi}_{n}(\boldsymbol{\tilde{x}}^{(t)}) + \frac{1}{\epsilon_{0}} (\boldsymbol{\tilde{x}}^{(t)} - \boldsymbol{x}),
\end{equation*}
where the subgradient of $\tilde{\varphi}_{n}(\boldsymbol{x}) = \max_{1\leq j \leq n} \{ f_{j} (\boldsymbol{x}) \} = \max_{1\leq j \leq n} \{ \langle \boldsymbol{x}, \tilde{\boldsymbol{x}}_{j} \rangle - \psi_{j}  \}$ is
\begin{equation*}
	\partial \tilde{\varphi}_{n}(\boldsymbol{x}) = \text{convex} \bigcup \{\partial f_{j} (\boldsymbol{x}) | f_{j} (\boldsymbol{x}) = \tilde{\varphi}_{n}(\boldsymbol{x})\}.
\end{equation*}
The stopping rule consists in satisfying the three conditions: (1) $h(\boldsymbol{\tilde{x}}^{(t)}) < rtol_{1}$; (2) $h(\boldsymbol{\tilde{x}}^{(t)}) < h(\boldsymbol{\tilde{x}}^{(t+1)})$, and (3) $|\partial h(\boldsymbol{\tilde{x}}^{(t)})| < rtol_{2}$; where $rtol1, rtol2$ are the tolerances given as input. 

\subsection{Computational cost}
The computational proposal in this section allows to compute the smoothest interpolation of the OT map in a lower theoretical running time than existing proposals.
Let us consider the construction of $T_{s}: \mathcal{X}_{s} \rightarrow \tilde{\mathcal{X}}_{s}$ for a given set $\mathcal{X}_{s}$, for each $s \in \{0, 1\}$.
Subsequently, once a new incoming sample has been divided into the two $m_s$-subsamples drawn from $\mu_{s}$, and such that $m_s \leq n_{s}$, we may need to provide the repaired version of each point. 
The naive approach, as discussed in the Introduction, involves computing a new OT map 
between the augmented $(n_{s} + m_s)-$samples and their associated Wasserstein barycenter. 
However, this would incur in a high computational time cost of $\mathcal{O}((n_{s} + m_s)^{3})$. 
In contrast, the interpolation methodology for $T_{s}$ proposed in this paper 
reduces the problem to solving $m_s$ optimization problems as in Eq. \eqref{eq:11} for each new point. 
Since the objective function $h$ presented in Eq. \eqref{eq:13} is Lipschitz with constant $\max_{1 \leq i \leq n} \Vert \boldsymbol{x}^{s}_{i} \Vert + \epsilon_{0}^{-1}$ and $\frac{1}{\epsilon_{0}}$ - strongly convex (claim 14.10 of \cite{10.5555/2621980}), an $\epsilon_{0}$ - optimal solution can be reached in $\mathcal{O}(\epsilon^{-1})$ epochs using Algorithm \ref{alg:stochastic}. 
Furthermore, evaluating $\partial \tilde{\varphi}_{n}$ at each iteration $t$ have a operational $n_{s}$ - cost. 
To sum up, computing the interpolation with precision $\epsilon$ of an $m_s$-sample has a computational complexity of order $\mathcal{O}(m n_{s} \epsilon^{-1}) + \mathcal{O}(\sqrt{n_{s}} n_{s}^{2} log(n_{s} C))$, where the first term corresponds to the Section \ref{sec:Moreau-Yoshida} and the second one to the Section \ref{sec:combinatorial_optimization}.
Refer to Table \ref{table:comp_costs} (Appendix \ref{app-add-plots}) for detailed information on computational costs.

\section{Simulations}
\label{sec:numericalexperiments}
\par
We evaluated our extended total repair method on simulated data, leading to some practical suggestions. Indeed, an improvement of the procedure has been identified in cases of low density of points in the support under interpolation, which is very beneficial for use with real data and will be addressed in  Section \ref{sec:discussion_implementation}.
\par
The first experiment ($\mathcal{E}$1) focused on applying the extended total repair 
in two different scenarios: ($\mathcal{E}$1-A) when there is an equal count of privileged and unprivileged individuals, and ($\mathcal{E}$1-B) in the general case. Furthermore, the repair is integrated within a 10-fold CV in order to train algorithms of different types and test the improvement in fairness metrics, particularly the DI. Lastly, the second experiment ($\mathcal{E}$2) consisted on applying the extended total repair in an online scenario in general dimension, when new biased data arrives and it is necessary to preprocess it according to the previous transformation. 
The generated datasets and the reproduction codes for this part can be found at \href{https://github.com/emartindedi}{https://github.com/emartindedi}.

\begin{itemize}

	\item[($\mathcal{E}$1)]\textit{ One-dimensional approach in two configurations: $n_{0} = n_{1}$ and $n_{0} \neq n_{1}$.} We generated ($\mathcal{E}$1-A) $n_{0} = n_{1} = 200$, and ($\mathcal{E}$1-B)  $n_{0} = 200$,  $n_{1} = 300$, samples from two multivariate normal distributions on $\mathbb{R}^{5}$; with means $(3, 3, 2, 2.5, 3.5)$ and $(4, 4, 3, 3.5, 4.5)$, respectively, and equal covariance matrix  $\text{diagonal}(1, 1, 0.5, 0.5, 1)$. Then, in order to introduce some bias in the dataset $X$, we obtained the classification label $Y$ with higher probability of success for the class $S = 1$. This is, $\pi_{s}(x) = \frac{1}{1 + e^{- X \beta_{s}}}$ with $\beta_{0} = (1, -1, -0.5, 1, -1, 1)$ and $\beta_{1} = (1, -0.4, 1, -1, 1, -0.5)$.
    \par
    The methodology has been applied in a validation set approach.  The idea is to apply the methodology in a setting where the discrete transport map gives us the repair version of the training data and the continuous approximation gives the repair features of the test data. The repair process has been only applied to the first variable, that is why we called one dimensional approach even for different characteristics. Specifically, for every fold $k$, the training data $(k-1) \cdot 100/k \%$ has been modified according to the discrete OT map, while the test data $100/k \%$ was subject to modification through the interpolation function. The results of the interpolation for one of the folds are shown in Figure \ref{im:1}. 
    The training sample is depicted by group, $S=0$ in blue and $S=1$ in green. Then, the repaired interpolated values for the test set have been added in colour purple and orange, respectively for $S=0$ and $S=1$.
    As noted in the works of \cite{barrainkua2024uncertainty, aghbalou2023bias, dwork2015reusable}, there may be a strong variability when computing the disparate impact of different subsamples of the data . Therefore, we performed 10-fold CV for three different algorithms: logistic regression (LR), decision trees (DT) and gradient boosting (GB). The results are compared against the biased measured in the data with their respective theoretical confidence interval of the DI.
    The obtained results for both configurations ($\mathcal{E}$1-A) (upper) and ($\mathcal{E}$1-B) (lower) are shown in Figure \ref{im:exp1}. We can observe that with the model GB we obtained the best trade-off between precision and non discrimination in the predictions.  For the exact values, we refer to Table \ref{table:exp2}. 

\begin{figure}[h!]
         \centering
         \includegraphics[width=0.6\textwidth]{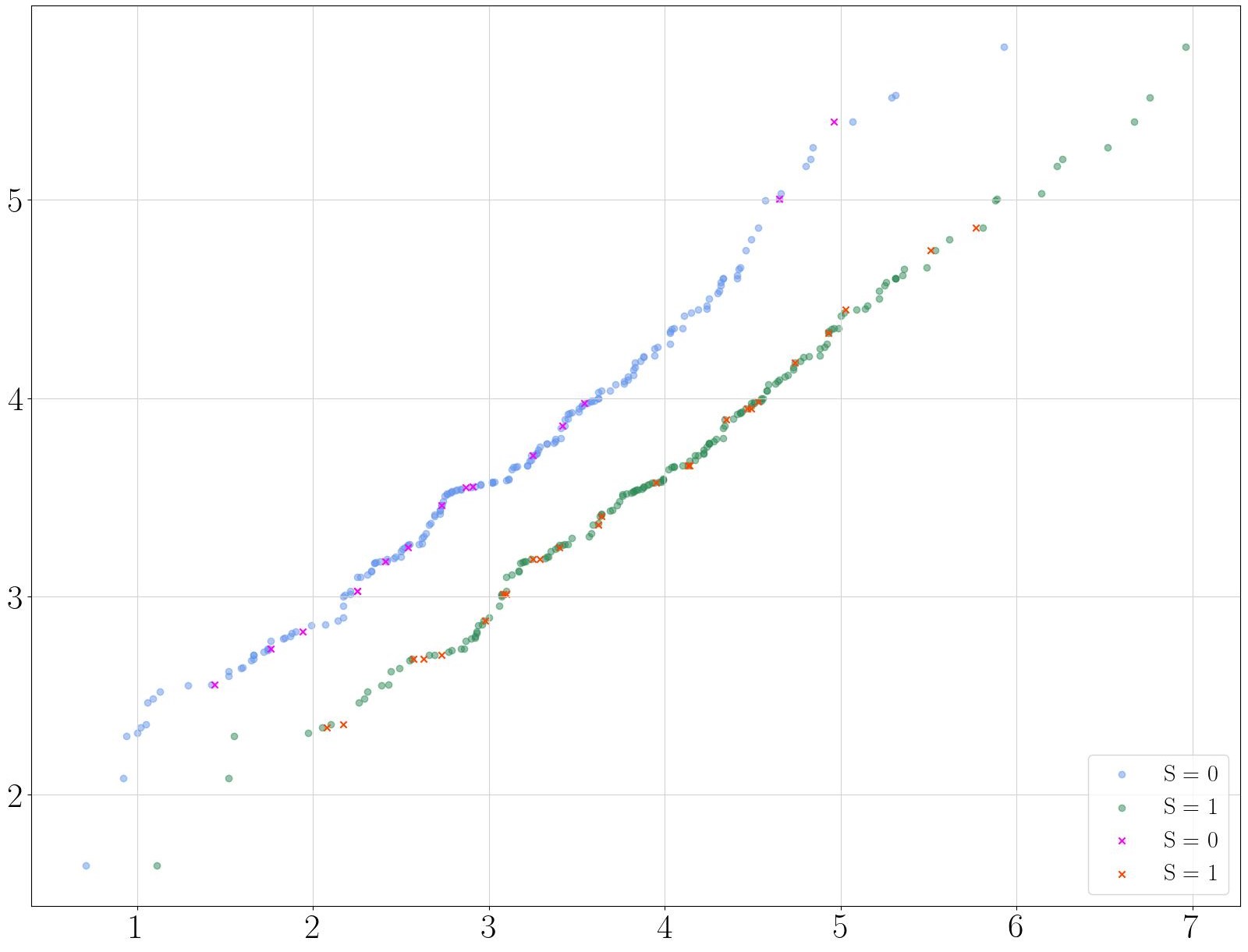}
         \caption{Repaired values of the training set $\mathcal{X}_s^{train}$ (blue and green dots) and test set $\mathcal{X}_s^{test}$ (purple and orange crosses), given by $T_s$ and its interpolation $\Bar{T}_{s}$, respectively. The x-axis shows the original values and the y-axis shows the transformed values.}
    \label{im:1}
\end{figure}

    \begin{figure}[h!]
         \begin{subfigure}[b]{0.49\textwidth}
             \includegraphics[width=.49\textwidth]{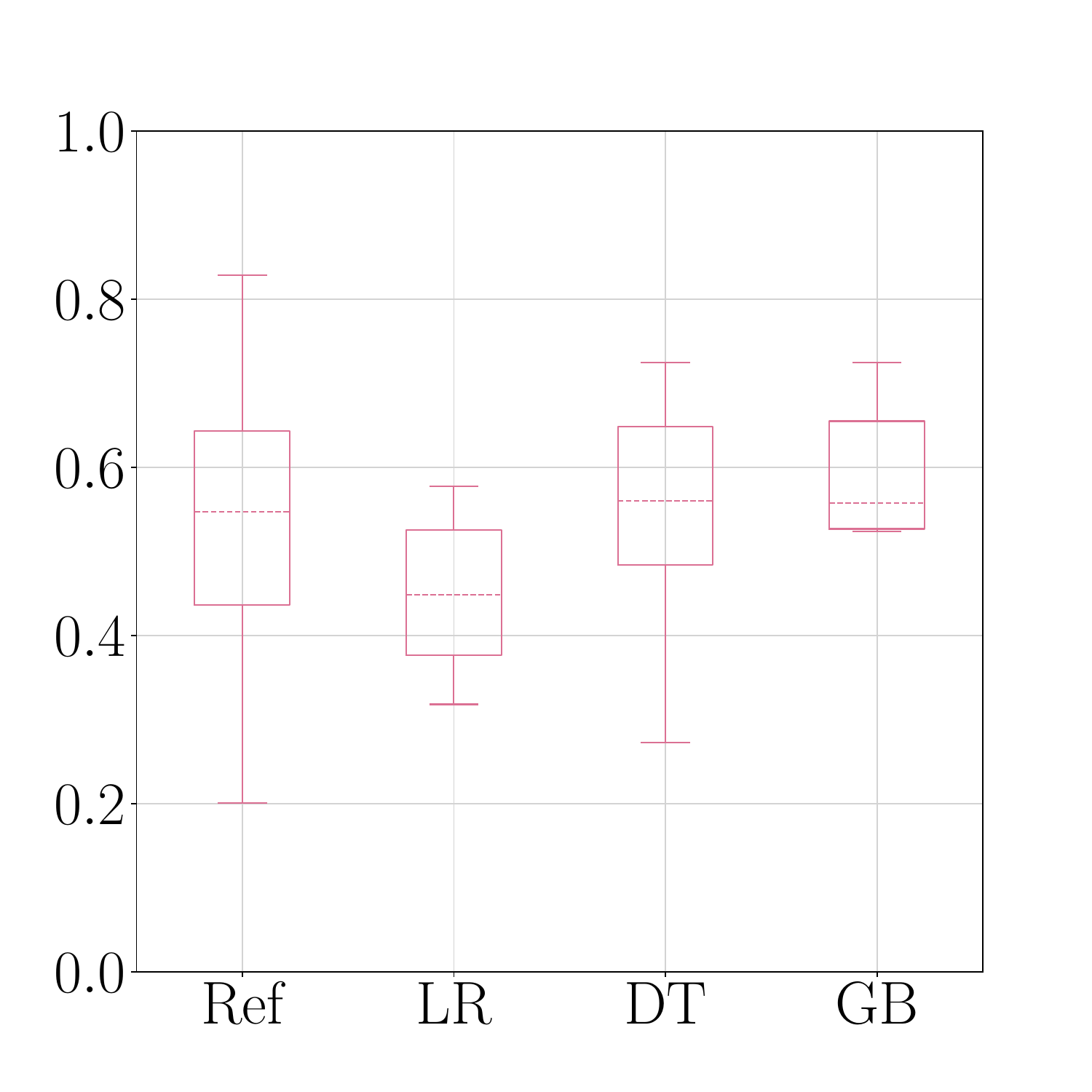}
            \includegraphics[width=.49\textwidth]{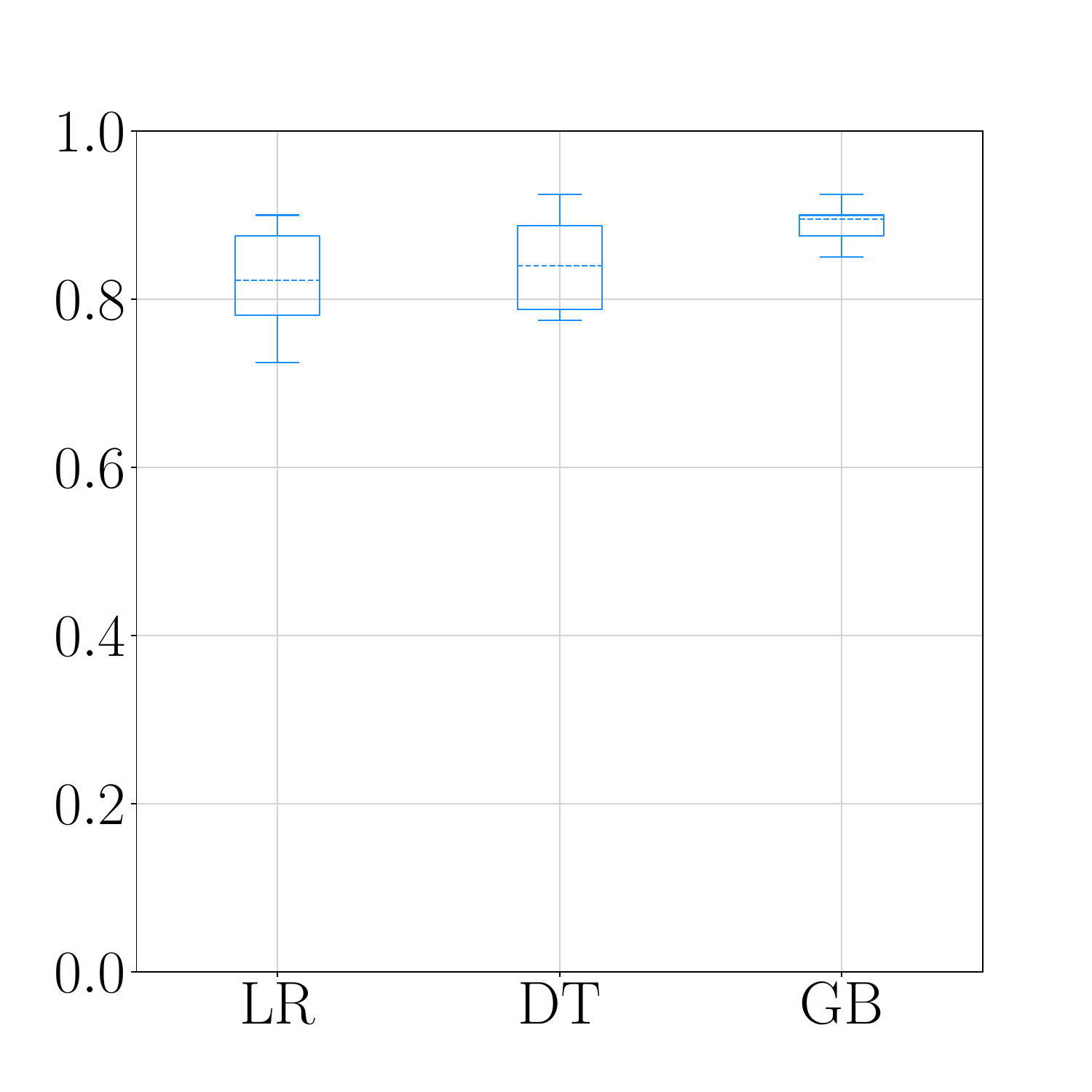}
             \caption{($\mathcal{E}$1-A) with the original dataset.}
             \label{im:2a}
         \end{subfigure}
         \begin{subfigure}[b]{0.49\textwidth}
             \includegraphics[width=.49\textwidth]{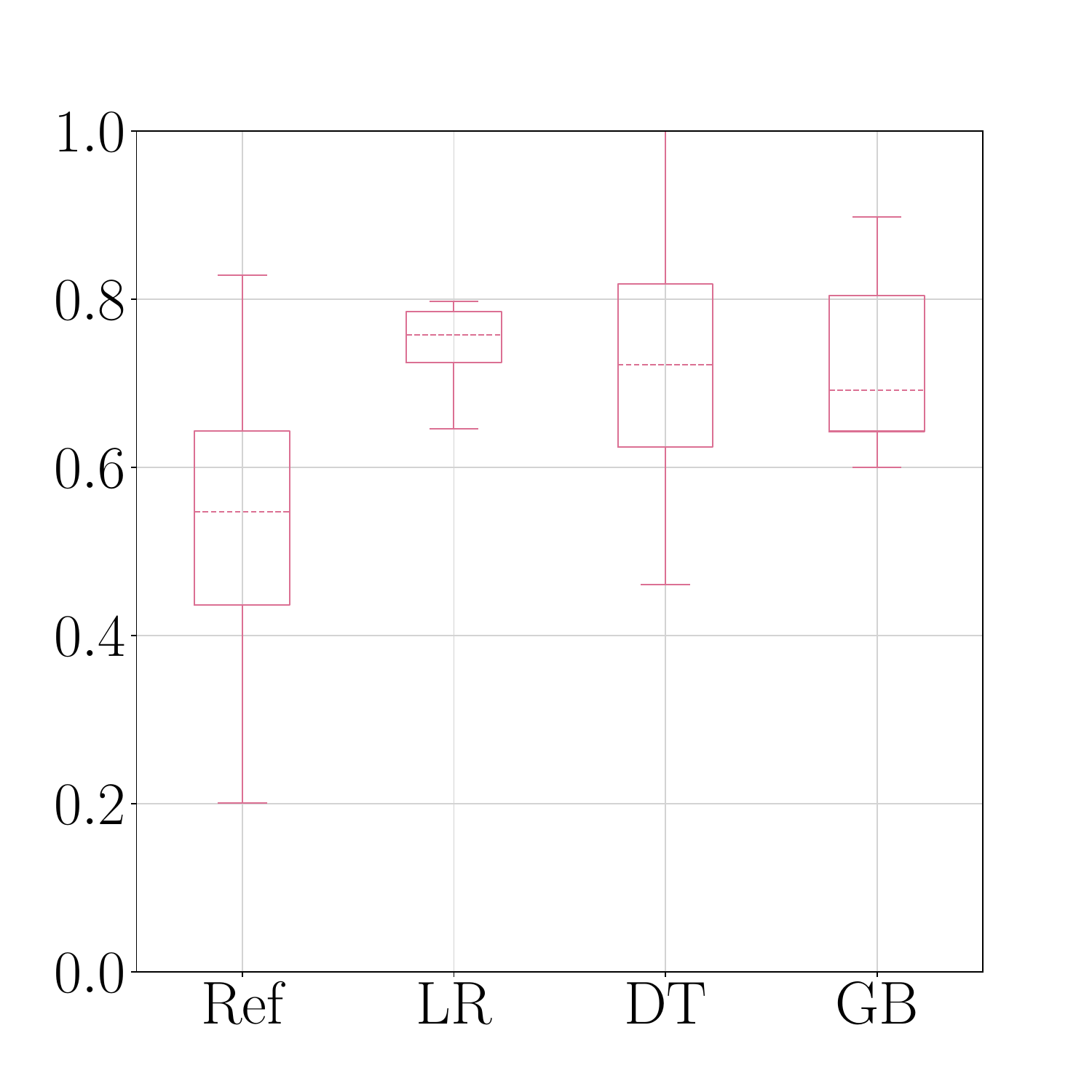}
            \includegraphics[width=.49\textwidth]{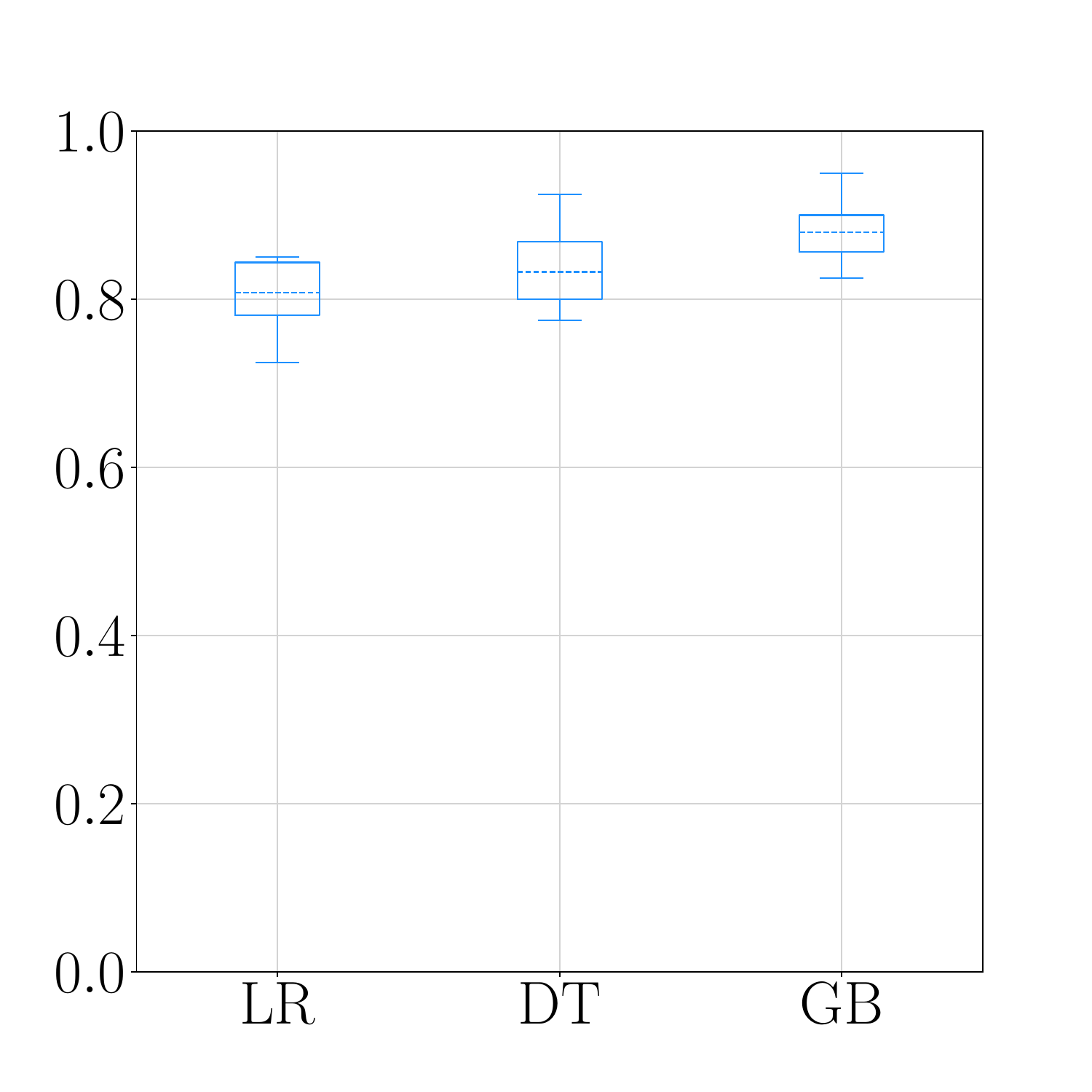}
             \caption{($\mathcal{E}$1-A) with the modified dataset.}
             \label{im:2b}
         \end{subfigure}
        \label{im:exp2a_boxplots}
        \begin{subfigure}[b]{0.49\textwidth}
             \includegraphics[width=.49\textwidth]{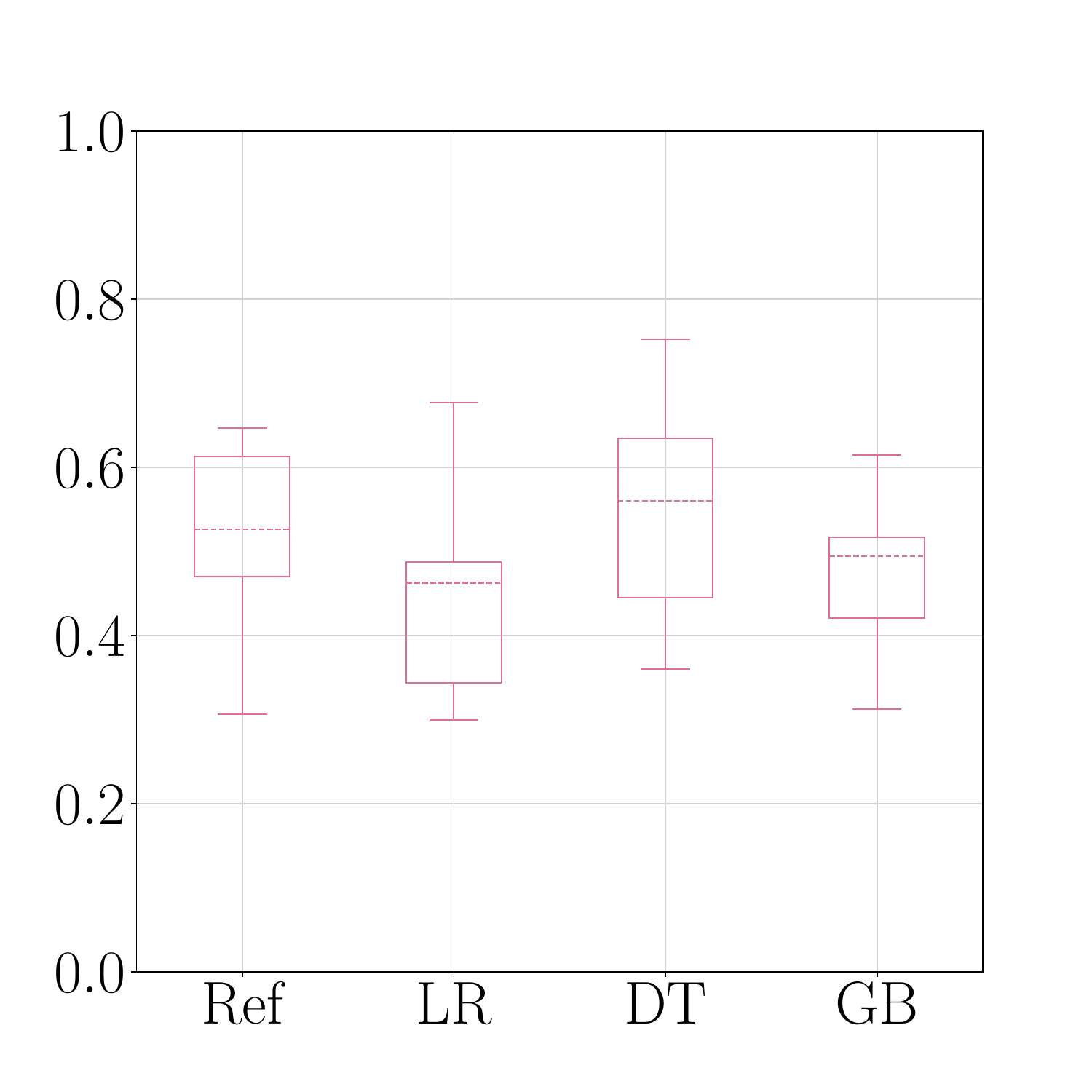}
            \includegraphics[width=.49\textwidth]{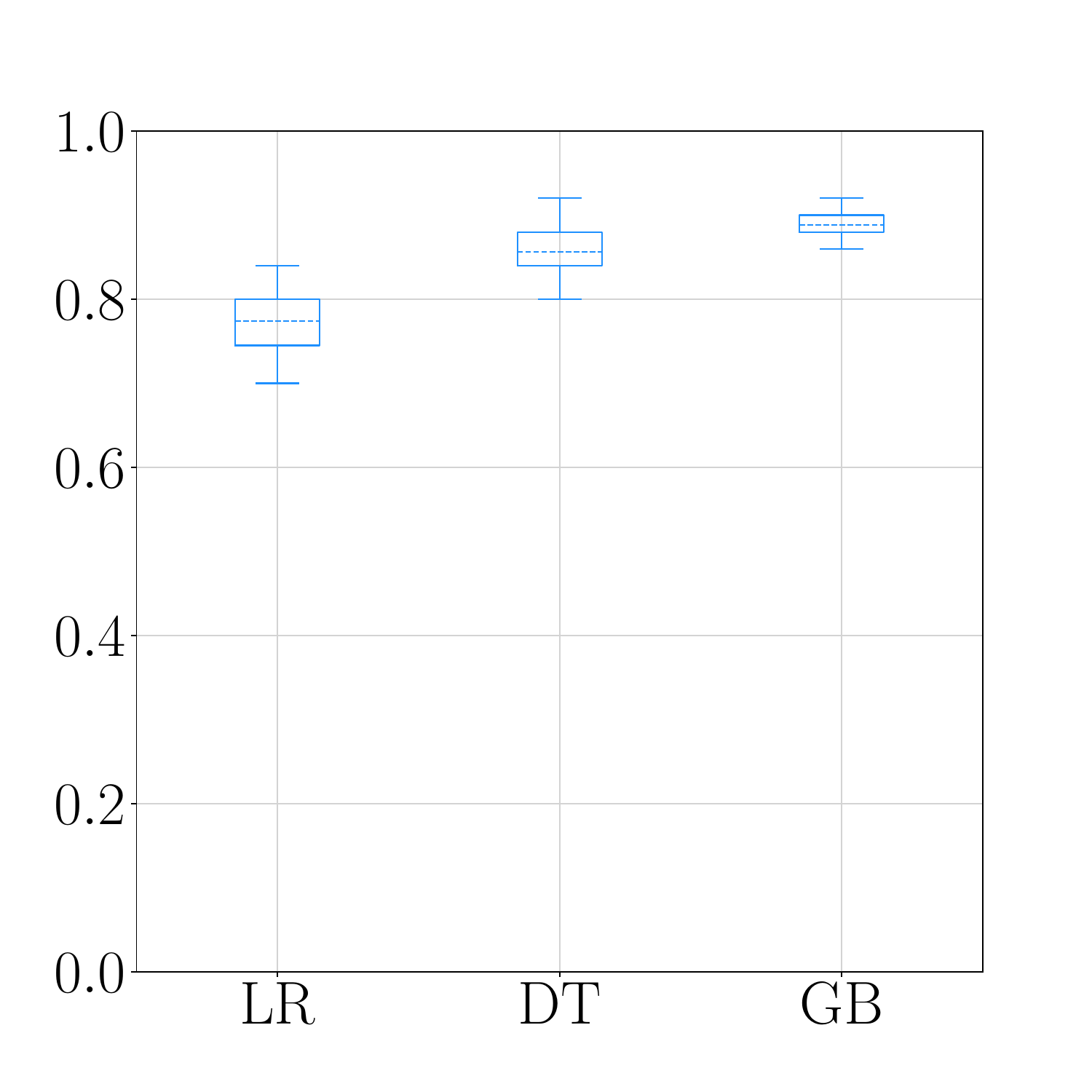}
             \caption{($\mathcal{E}$1-B) using the original dataset.}
             \label{im:2c}
         \end{subfigure}
         \begin{subfigure}[b]{0.49\textwidth}
             \includegraphics[width=.49\textwidth]{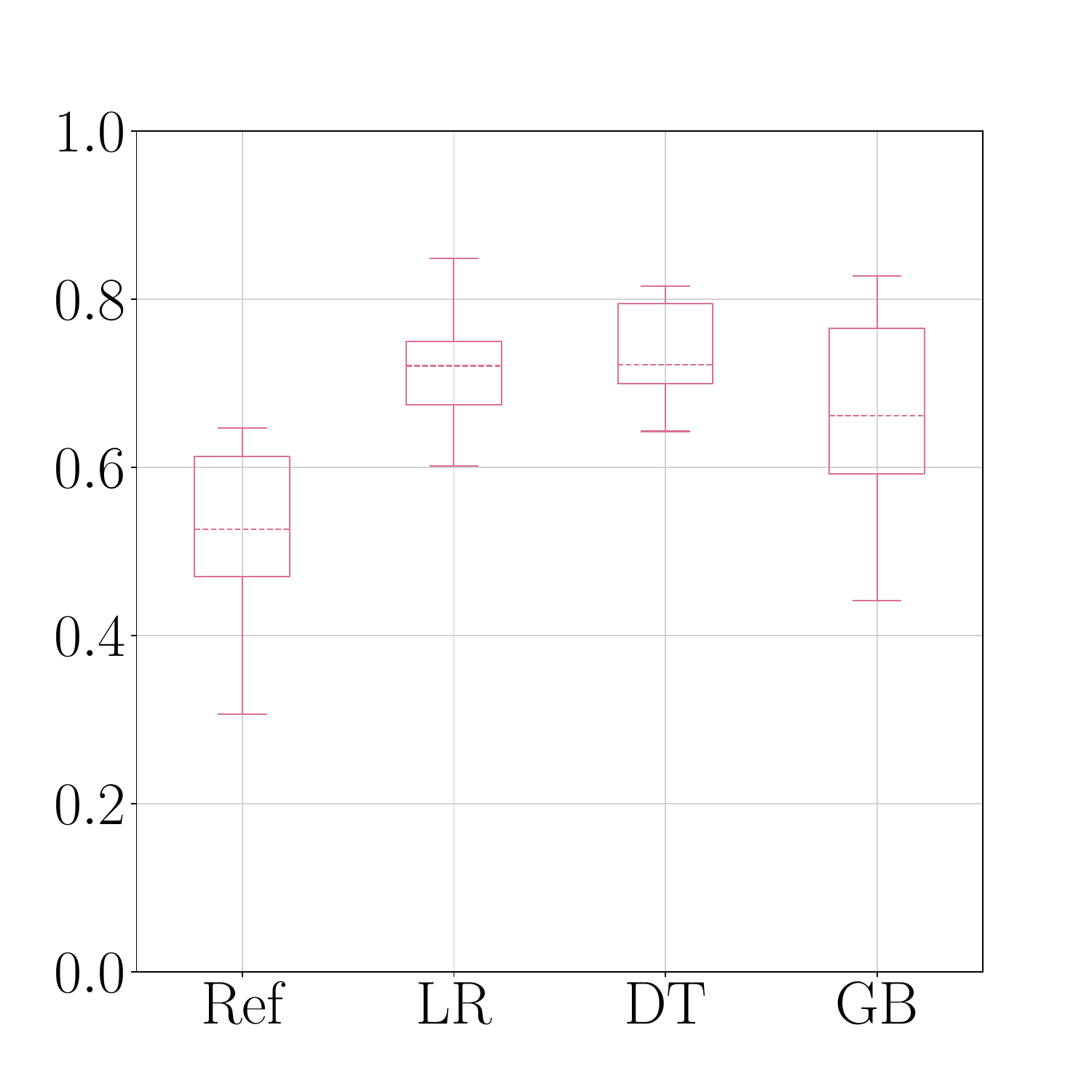}
         	\includegraphics[width=.49\textwidth]{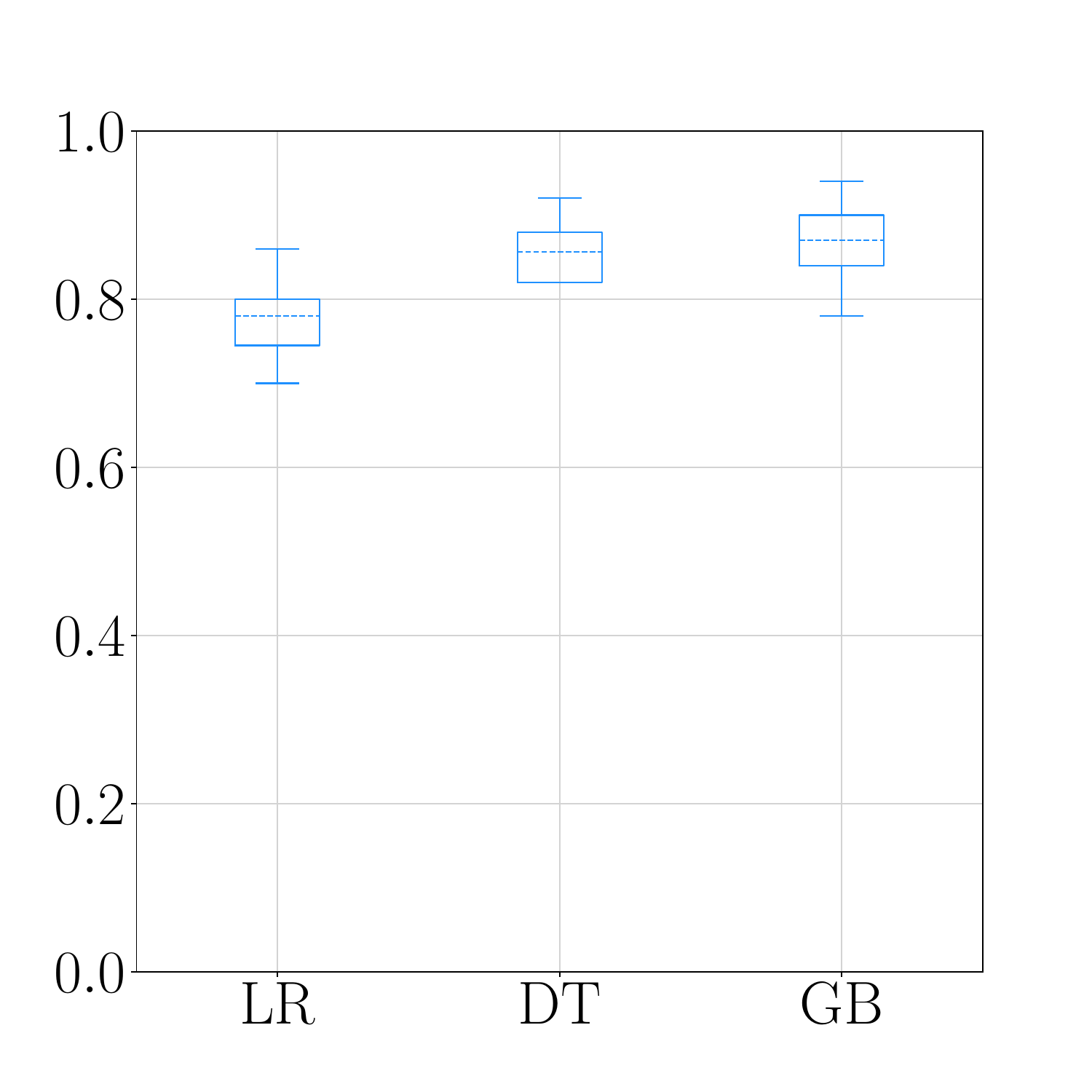}
                \caption{($\mathcal{E}$1-B) using the modified dataset.}
             \label{im:2d}
         \end{subfigure}
        \caption{Results of experiment ($\mathcal{E}$1): Fairness (disparate impact - pink) and performance (accuracy - blue) metrics measured in a $10$-fold CV scenario. Upper row (a-b) corresponds to configuration A, while bottom row (c-d) to B. The dashed lines are the average value.}
        \label{im:exp1}
    \end{figure}

    \item[($\mathcal{E}$2)]\textit{$d$-dimensional approach in an online scenario.} We generated $n_{0} = 200$ and $n_{1} = 200$ samples from two different multivariate normal distributions on $\mathbb{R}^{3}$ representing a dataset obtained in an offline situation. However, in a ML production environment new data appears. Hence,  an opportunity to generate novel predictions through retraining the AI model with the new data is shown. This variables need to be consistently modified (repair) according to the previous transformation. Subsequently, we generate an additional $m_{0}= 40$ and $m_{1} = 40$ samples from the two sub-populations standing for online data, maintaining the same distribution as the earlier reference data. The Wasserstein barycenter (S=
    0 purple, S=1 black) between the offline subsamples  (S=
    0 blue, S=1 green) is depicted in Figure \ref{im:exp3} (left). Then, the interpolated values of the OT maps to such a barycenter are computed for the incoming online data in Figure \ref{im:exp3} (right), with the same colour coding by groups. From a visual standpoint, the discrete Wasserstein barycenter provides a natural extension of the notion of averaging points to the notion of averaging point clouds.    
    \begin{figure}[h!]
        \centering
            \includegraphics[width=.4\textwidth]{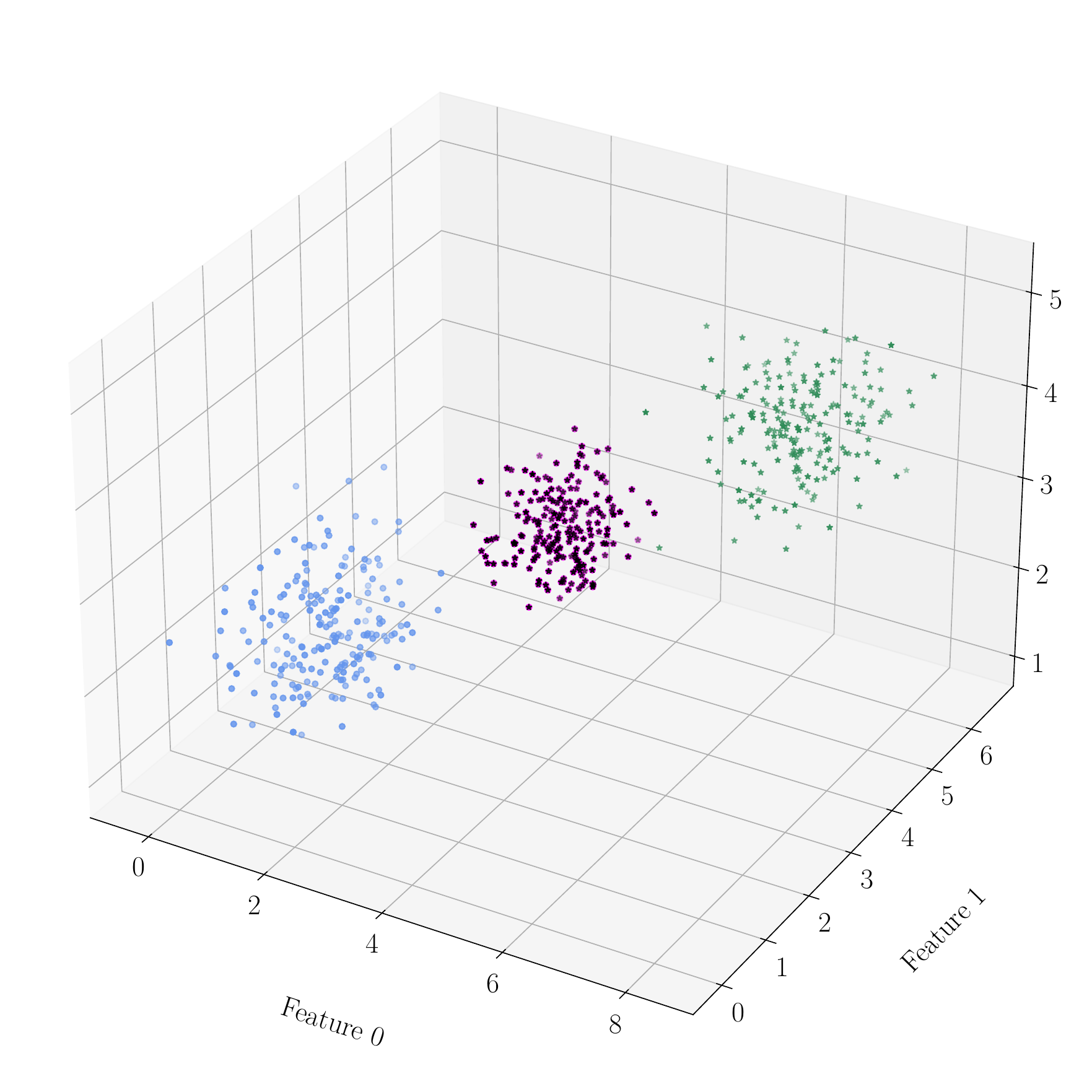}
            \includegraphics[width=.4\textwidth]{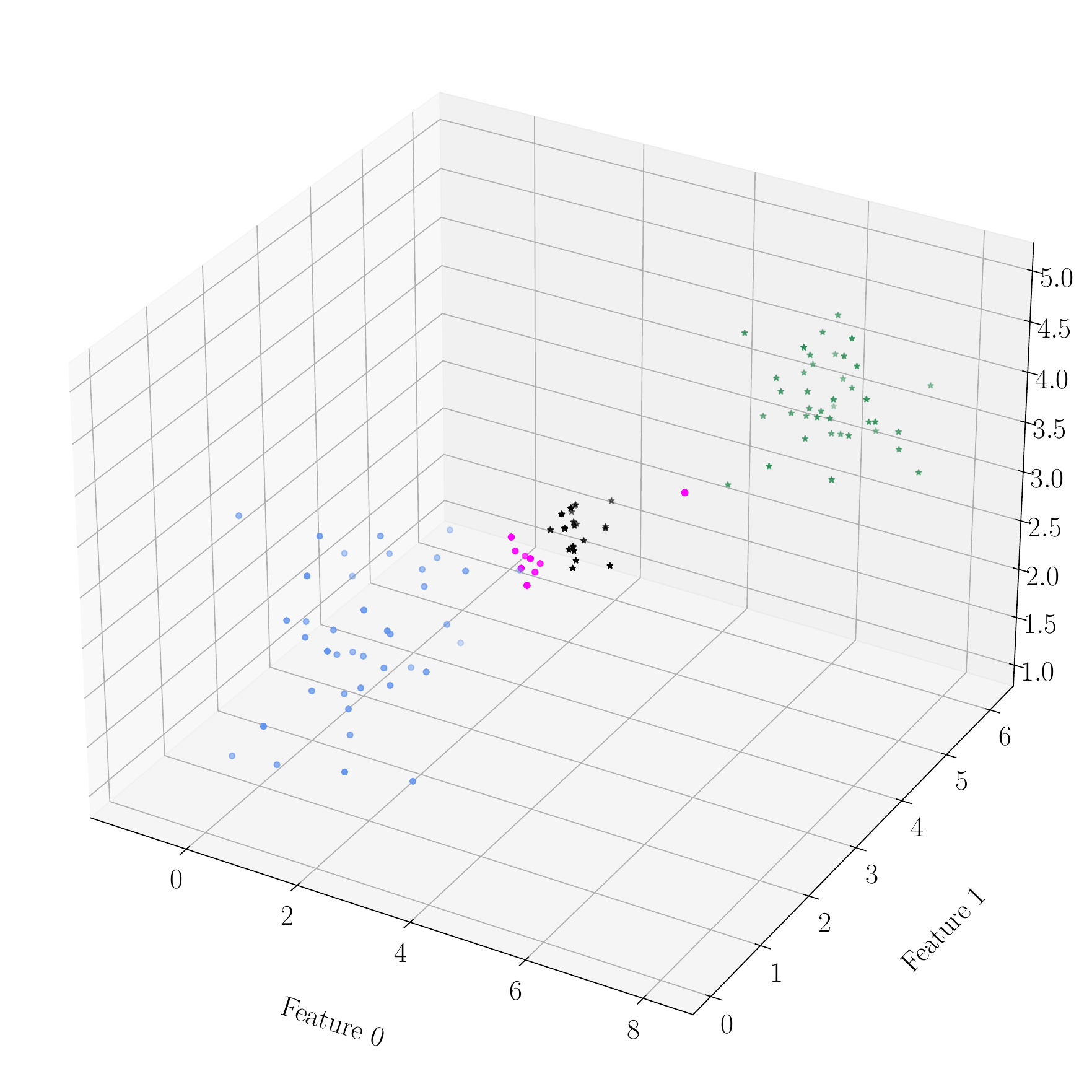}
            \caption{Experiment ($\mathcal{E}$2): For both groups, $ S = 0$ (blue) and $S = 1$ (green), the modified version is illustrated, colored in purple and black respectively, for the offline data (left) and the online data (right).}
            \label{im:exp3}
    \end{figure}

\end{itemize}

\section{Real data application}
\label{sec:applications}
\par
In this section, we first explain how to adapt the methodology to real data to improve the results (Section \ref{sec:discussion_implementation}). Then, we describe the dataset employed in the evaluation of the proposed methodology (Section \ref{sec:datadescription}). 
After that, we define the experimental setup, first via the classical benchmark (without fairness constraints), and  secondly via our extended total repair proposal, providing
 an analysis of results (Section \ref{sec:datadescription}). 

\subsection{Discussion of the implementation of the extended total repair with real data}
\label{sec:discussion_implementation}
The bias mitigation methodology for fair classifications introduced in Section \ref{sec:theo_extended} focuses on removing biases inherent in datasets, particularly those associated with sensitive attributes (e.g. age, gender), through a pre-processing technique based on OT theory. Essentially, it consists of calculating the transport plan to the Wasserstein barycentre between the non-protected characteristics $\mathbf{X}$ of each demographic group $S$, and then constructing a smooth interpolation that guarantees cyclic monotonicity. The latter entails a number of computational challenges that have been addressed in the previous Section \ref{sec:computational_aspects}. Still, some additional issues need to be clarified when putting it into practice with real data, which are discussed in this section and will be taken into account to obtain the results of Section \ref{sec:applications} . Specifically, depending on the size of the new sub-samples to be repaired, different alternatives are considered. As a result, we propose a global procedure, outlined in the pseudo-code given in Procedure \ref{pipe:exp_setup}, that integrates: (i) the repair of the original data, (ii) three different versions of the calculation of the interpolation, and (iii) the learning of a classifier in a cross-validation scheme that also allows to reliably measure the accuracy and fairness of the trained rule. 
\par
Consider the dataset $D = \{(\boldsymbol{x}_{0}, s_0, y_{0}), ..., (\boldsymbol{x}_{N}, s_N, y_{N}) \text{ } |  \text{ } \boldsymbol{x}_{i} \in \mathcal{X}, s_{i} \in \{0,1\}, y_{i} \in \{0,1\} \}$ resulting from an initial pre-processing and cleaning step. Before explaining the procedure further, it should be noted that a prior analysis of the variables $\mathbf{X}$ is necessary, on the one hand, to select those that need to be transformed and, on the other hand, to identify those that should not be modified. In this task, quantitative fairness metrics are used (for e.g, DI) to test for differences between the two $S$ groups. We note that this question depends strongly on the context of the problem and the concrete data at hand. We suggest that this decision should involve a diverse group of experts from various fields, particularly sociology, who can provide a deeper understanding of the historical, cultural and systemic factors that might contribute to unfair predictions; as well as the field in which the algorithm in question is intended to be applied (e.g. if a decision is to be made on whether a patient receives certain medical treatment, the opinion of the medical staff in charge of the treatment should be taken into account).  Let us denote this selection of variables by $\mathbf{X}^{\star}$, which is given as input to Procedure \ref{pipe:exp_setup}. 

Analogously to the previous section, in order to apply the methodology to real data sets,
we propose to use three different algorithms: logistic regression (LR), decision trees (DT) and gradient boosting (GB).
The proposed methodology is integrated with a set validation approach by means of K-fold CV. In addition, to compare the results obtained using the extended total repair, it is useful to train the algorithms without taking into account fairness considerations. This is implemented as \textbf{Classical benchmark} in Procedure \ref{pipe:exp_setup} that obtains a classifier $g_{k}$ for each fold $k$, using the training data $X^{\text{train}}_{k}$ and performance and fairness metrics $L_{X^{\text{test}}_{k}}(g_{k})$ and $DI(g_{k}, X^{\text{train}}_{k}, S)$ respectively.
\par
Then, the \textbf{Extended Total Repair} is broadly outlined for each fold $k$.
The approach encompasses a dual-stage process wherein the discrete OT mapping $X^{\star \text{train}}_{\text{k}}$ to the data $\tilde{X}^{\star \text{train}}_{k}$ (Section \ref{sec:total-repair}) is initially computed, followed by the subsequent calculation of the interpolation function (Sections \ref{sec:theo_extended} and \ref{sec:computational_aspects}). This second phase can be implemented in three different ways depending on the precision of the continuous function needed.
In scenarios where the sizes of each group $n_{s} = | \mathcal{X}_{s} |, \ s \in \{0, 1\}$ are sufficiently large, using the initial approximation of the interpolation function suffices (\textbf{Option 1}). Hence, for a new point $\boldsymbol{x} \in \mathbb{R}^{d}$, the modified version will be $\Bar{T}_{s}(\boldsymbol{x}) = \nabla \tilde{\varphi}_{n}(\boldsymbol{x})$, where $\tilde{\varphi}_{n}$ is defined as in Eq. \eqref{eq:7}. Otherwise,
for small sizes, the regularized function $\Bar{T}_{s} = \nabla\varphi_{\epsilon_{0}}$ is computed  (\textbf{Option 2}). We notice that we could find a mixture of the two situations, i.e. cases where the support has a high density of points in some intervals and not in others.
Thus, we additionally propose an hybrid option of the two previous ones, which will consists in adding the regularized step only in the feature space where the empirical distributions of the subgroups have higher density than a threshold (\textbf{Option 3}). 
This flexibility allows for adaptation of the approach by incorporating the regularization step only as required.

\setcounter{AlgoLine}{10}
\SetKwBlock{Begin}{}{end}

\renewcommand{\algorithmcfname}{Procedure}%
\begin{algorithm}[ht]
\SetAlgoNoLine
\caption{Pre-processing bias mitigation methodology}
\label{pipe:exp_setup}

\SetKwBlock{Begin}{Begin}{}

\KwInput{

\Indp
Dataset $D = \{(\boldsymbol{x}_{0}, y_{0}), ..., (\boldsymbol{x}_{N}, y_{N}) \}$, where $\boldsymbol{x}_{i} \in \mathcal{X}, y_{i} \in \mathcal{Y}$\;
Binary protected attribute from the demographic characteristics (e.g. $ S = ``\text{age}"$)\;
Numeric variables of interest for the bias mitigation pre-processing method $X^{\star}$\;
\Indm

}

\KwOutput{Prediction rule $g : \mathcal{X} \rightarrow \mathcal{Y} \in A(X^{\text{train}})$ with a trade-off between fair and performance metrics.}

Pre-processing step of the dataset $D$ \;
Supervised learning classification algorithm $A \in \{\mathcal{H}_{\text{linear predictors}}, \mathcal{H}_{\text{decision tree}}, \mathcal{H}_{\text{boosting}}\}$\;

\textbf{Classical benchmark}

\Indp
Create a $K-$ fold cross validation according to the validation set approach\;
\For{each fold $k$}{
$g_{k} = A(X^{\text{train}}_{k})$\;
Obtain the error of the prediction rule $L_{X^{\text{test}}_{k}}(g_{k})$\;
Obtain the disparate impact of the prediction rule $DI(g_{k}, X^{\text{train}}_{k}, S)$\;

}

Report performance metrics $\text{error} = \frac{1}{K} \sum_{k = 1}^{K} L_{X^{\text{test}}_{k}}(g_{k}) $ \;
Report fairness metrics $\text{DI average} = \frac{1}{K} \sum_{k = 1}^{K} DI(g_{k}, X^{\text{train}}_{k}, S) $ \;

\Indm



\textbf{Fairness proposal: Extended Total Repair}

\Indp
Create the $k-$ fold cross validation according to the validation set approach\;
\For{each fold}{

Map the training data $X^{\star \text{train}}_{\text{k}}$ to the data $\tilde{X}^{\star \text{train}}_{k}$ whose distribution is the Wasserstein barycenter of the two empirical laws $\mathcal{L}(X | S = s)$ for $s = 0,1 $ respectively \;

\textbf{Option 1: no regularization} (suitable for large sample sizes)

\Indp
Construct the function $\Bar{T}_{s} = \nabla \tilde{\varphi}_{n}$\;
Map the test data $X^{\star \text{test}}_{k}$ to the data $\tilde{X}^{\star \text{test}}_{k}$ with the function $\Bar{T}_{s}$ \;
\Indm

\textbf{Option 2: add a regularization step}

\Indp
Construct the function $\Bar{T}_{s} = \nabla\varphi_{\epsilon_{0}}$ \;
Map the test data $X^{\star \text{test}}_{k}$ to the data $\tilde{X}^{\star \text{test}}_{k}$ with the function $\Bar{T}_{s}$ \;
\Indm

\textbf{Option 3: a hybrid version}

\Indp

In the intervals characterized by a higher density of data points within the empirical distribution, we will apply the first option and the second one otherwise.

\Indm

Construct the $\tilde{X}^{\text{train}}_{k}$ by merging the original variables ($X^{\text{train}}_{k} \setminus X^{\star \text{train}}_{\text{k}}$) and the transformed variables of interest ($\tilde{X}^{\star \text{train}}_{k}$)\;

Construct the $\tilde{X}^{\text{test}}_{k}$ by merging the original variables ($X^{\text{test}}_{k} \setminus X^{\star \text{test}}_{\text{k}}$) and the transformed variables of interest ($\tilde{X}^{\star \text{test}}_{k}$)\;

$g_{k} = A(\tilde{X}^{\text{train}}_{k})$\;
Obtain the error of the prediction rule $L_{\tilde{X}^{\text{test}}_{k}}(g_{k})$\;
Obtain the disparate impact of the prediction rule $DI(g_{k}, \tilde{X}^{\text{test}}_{k}, S)$\;

}

\Indm

\end{algorithm}

\subsection{Description of the data and problem statement}
\label{sec:datadescription}
\par
The German credit dataset \citep{misc_statlog_(german_credit_data)_144}, comprises records of individuals who hold bank accounts. This dataset serves the purpose of forecasting risk, specifically to assess whether  to extend credit to an individual or not is advisable. 
Specifically, it contains information about 1000 individuals, described as values of 22 features: 13 categorical and 8 numerical. 
The objective of this dataset is to accurately predict the customer's level of risk when granting a credit, taking into account factors such as the status of the existing checking account, credit amount or marital status. 
Notably, the dataset contains potential demographic attributes such as age or gender. 
Therefore, it has gained significant popularity within the algorithmic fairness community as a reliable benchmark for evaluating and comparing the effectiveness of various bias mitigation techniques.
\par
When assessing possible biases and quantifying group fairness, it is important to examine certain demographic information within the data. 
The purpose of this analysis is to enhance our understanding of the provided information and uncover potential sources of inequity. 
After performing exploratory analysis 
it becomes clear that the dataset suffers from an unbalanced distribution of the different groups based on the sensitive attributes such as gender or age. 
\par
Regarding gender, the composition of individuals is unbalanced in favor of males, constituting approximately $69\%$ of the population.
Besides, when we discretize the age into a binary categorization of "young" (comprising individuals aged $25$ or younger) and "senior" (comprising individuals aged over $25$), it can be seen that there is a pronounced under-representation of young individuals.
This pivotal point in the age  is chosen by the fact that the age threshold of $25$ years represents a demographic division where the probability of success (pertaining to credit extension) for an individual, conditioned upon their age group, is notably diminished (See Figure \ref{im:threshold} of \ref{app-add-plots}). There is an unbalaced situation again with 81\% of the people in th smaple are senior.
From an exploratory perspective, it is valuable to investigate the base rates (i.e., class label) within the protected variables (see Table \ref{tab:6}). 
Notice that the distribution of the label attribute is significantly unbalanced, with only $30\%$ of individuals having a "no credit" class label. 
Moreover, the rates of level of risk in males and females are notably different, and so they are among age groups (see Table \ref{tab:6}). 
As a result, these particular features appear to be potentially sensitive variables for our study.


\par
Seeking to evaluate the presence of bias in a dataset or in data-driven decisions, the statistical parity criteria will be the reference in this study, quantified through the so-called disparate impact (DI).
Recall that $DI(Y, S) = \frac{P(Y = 1 | S = 0)}{P(Y = 1 | S = 1)}$
can be empirically estimated as $\frac{n_{1,0}}{n_{0,0} + n_{1,0}} / \frac{n_{1,1}}{n_{0,1} + n_{1,1}},$ where $n_{i,j}$ is the number of observations such that $Y = i, S = j$.
Notice that this estimation could be unstable due to the unbalanced amount of individuals on each class group. 
The closer the DI index is to $1$, the weaker is the discrimination between groups. 
In Table \ref{table:di}, we have quantified the DI and their respective confidence intervals \citep{besse2018confidence} for the bias already present in the original dataset with the sensitive attributes gender and age. A wage gap is seen both between men and women, as well as between young people and seniors. This is beyond the objective of this paper, but it is an important issue to be considered in the whole repairing process. 
Finally, we also note that the extension of the fairness criterion, replacing the true label $Y$ with the algorithm's outcome $g(X)$, is named $DIA(g, X, S)$ and will be addressed in a later stage. 

\subsection{Experimental setup and analysis of results}
\label{sec:realdataresults}
\par
Analogously to previous section of experiments,  we propose three different state-of-the-art supervised learning models: Logistic Regression (LR), Decision Tree (DT) and Gradient Boosting (GB) in order to predict the level of risk when extending a credit to an individual based on the features described.
The selection of these algorithms stems from their ability to encompass different modeling approaches, each of them representing alternative learning paradigms.
It is important to note that the aforementioned modeling approaches do not incorporate any algorithmic fairness constraints throughout their modeling process. 
Consequently, they serve as reference solutions against which bias mitigation techniques can be evaluated and compared.
We present two different experiments with the German Credit dataset, referred to as ($\mathcal{R}$1) and ($\mathcal{R}$2), each of them following the approaches summarized in Procedure \ref{pipe:exp_setup} as a pseudocode.

\begin{figure}[h!]
            \begin{subfigure}[b]{0.36\textwidth}
                \includegraphics[width=\textwidth]{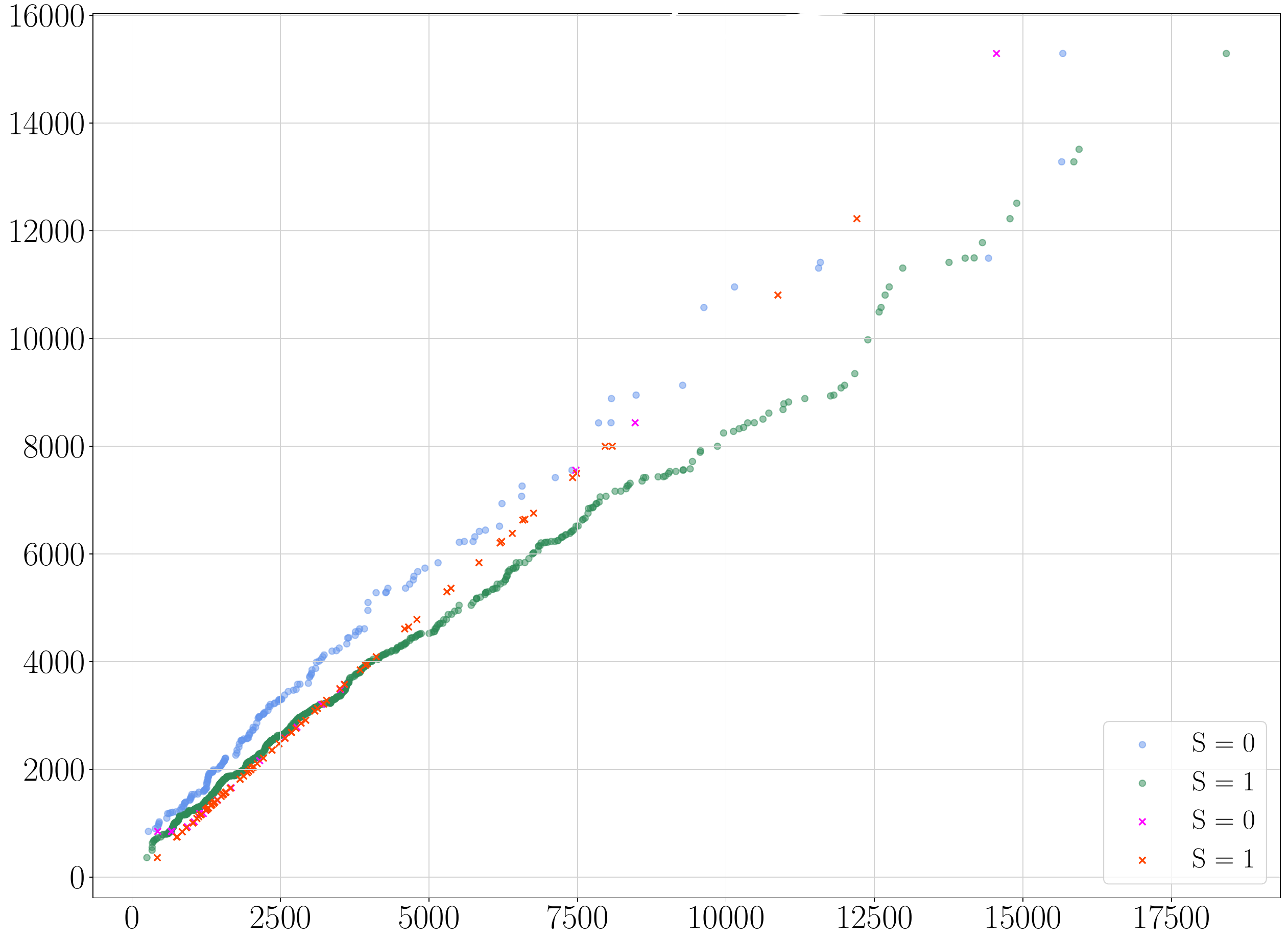}
            \caption{Original vs. repaired credit}
            \label{im:8a}
     	\end{subfigure}
     	\begin{subfigure}[b]{0.31\textwidth}
         	\includegraphics[width=\textwidth]{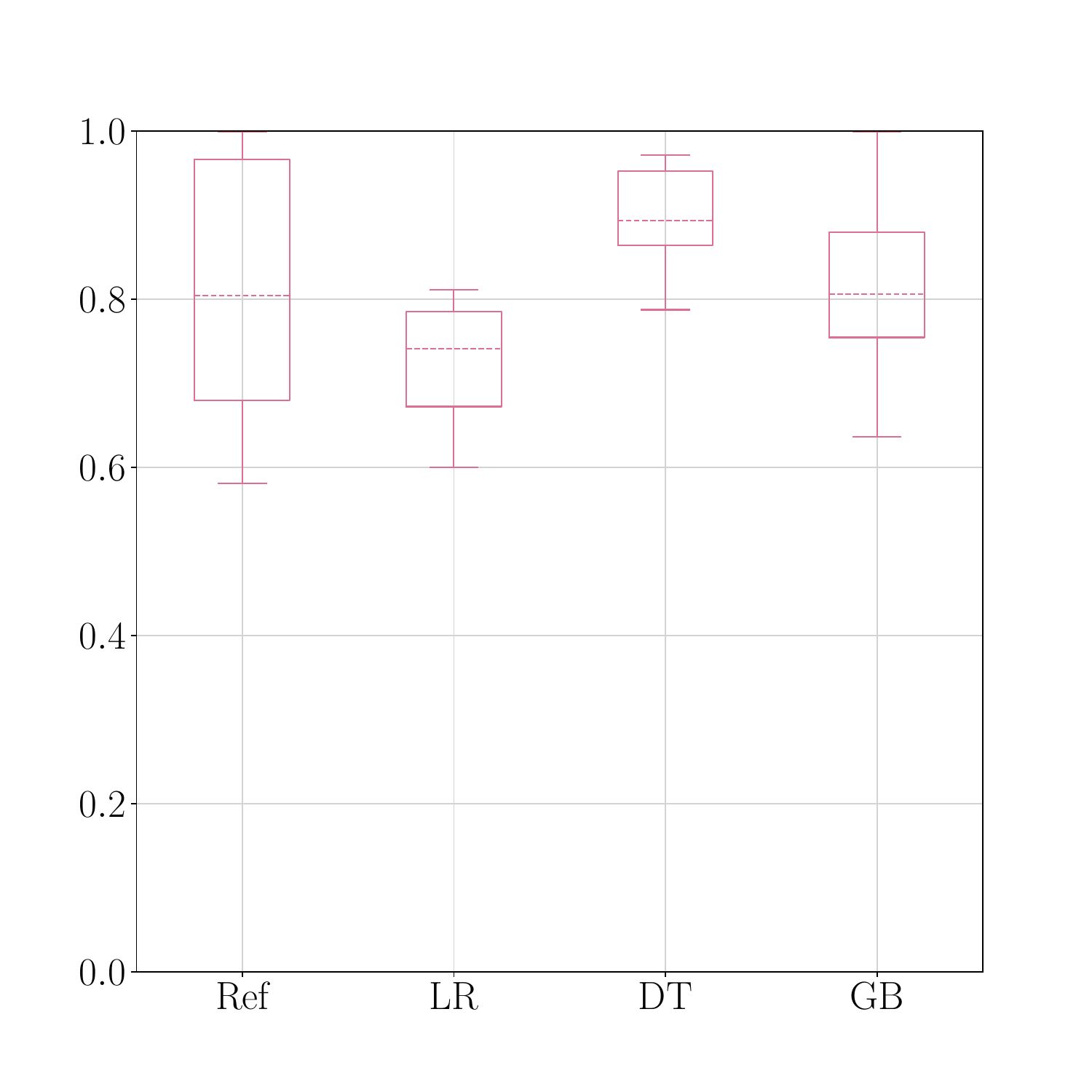}
            \caption{Disparate impact}
         	\label{im:8b}
     		\end{subfigure}
\begin{subfigure}[b]{0.31\textwidth}
\includegraphics[width=\textwidth]{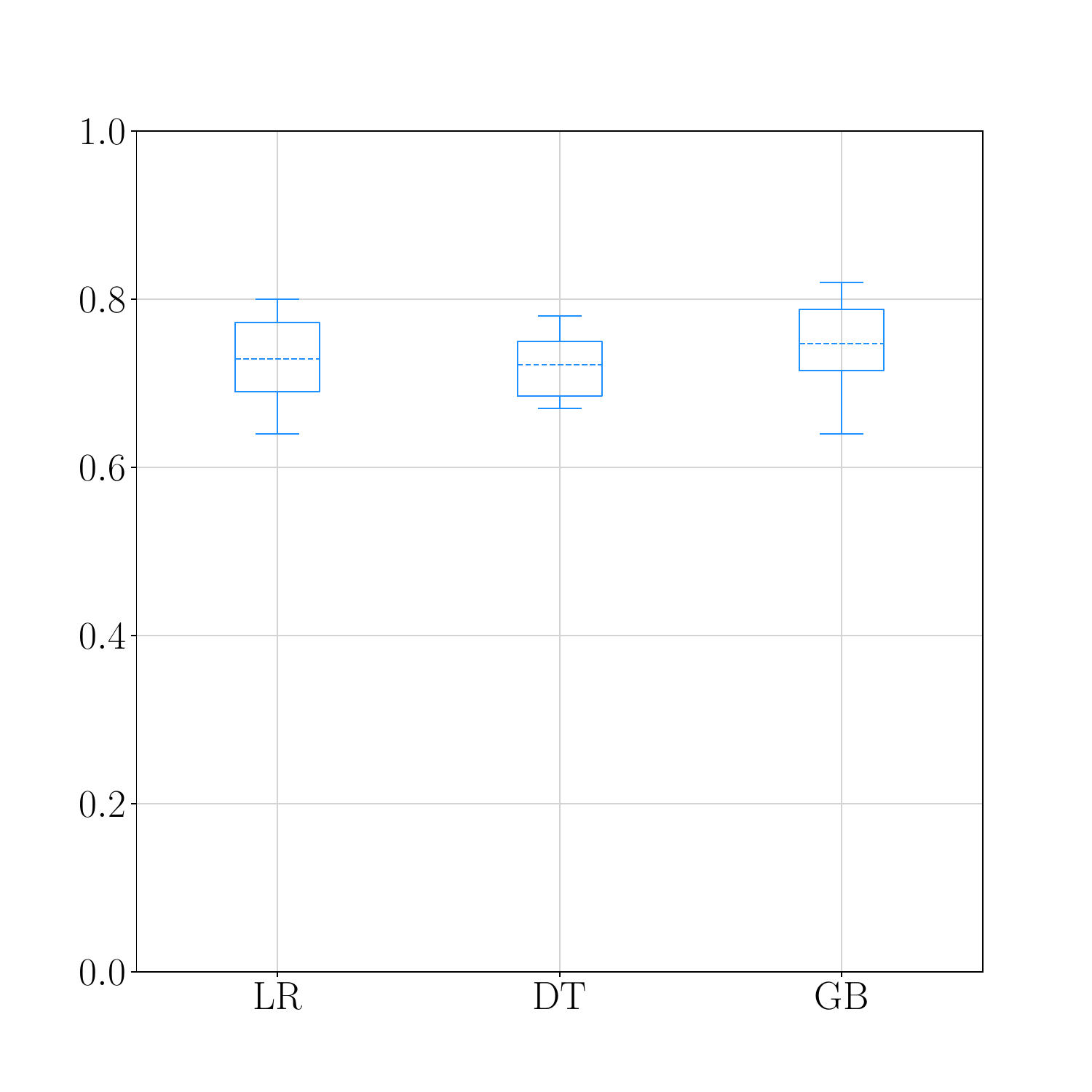}
         	\caption{Accuracy}
         	\label{im:8c}
     	\end{subfigure}

\begin{subfigure}[b]{0.36\textwidth}
    \includegraphics[width=\textwidth]{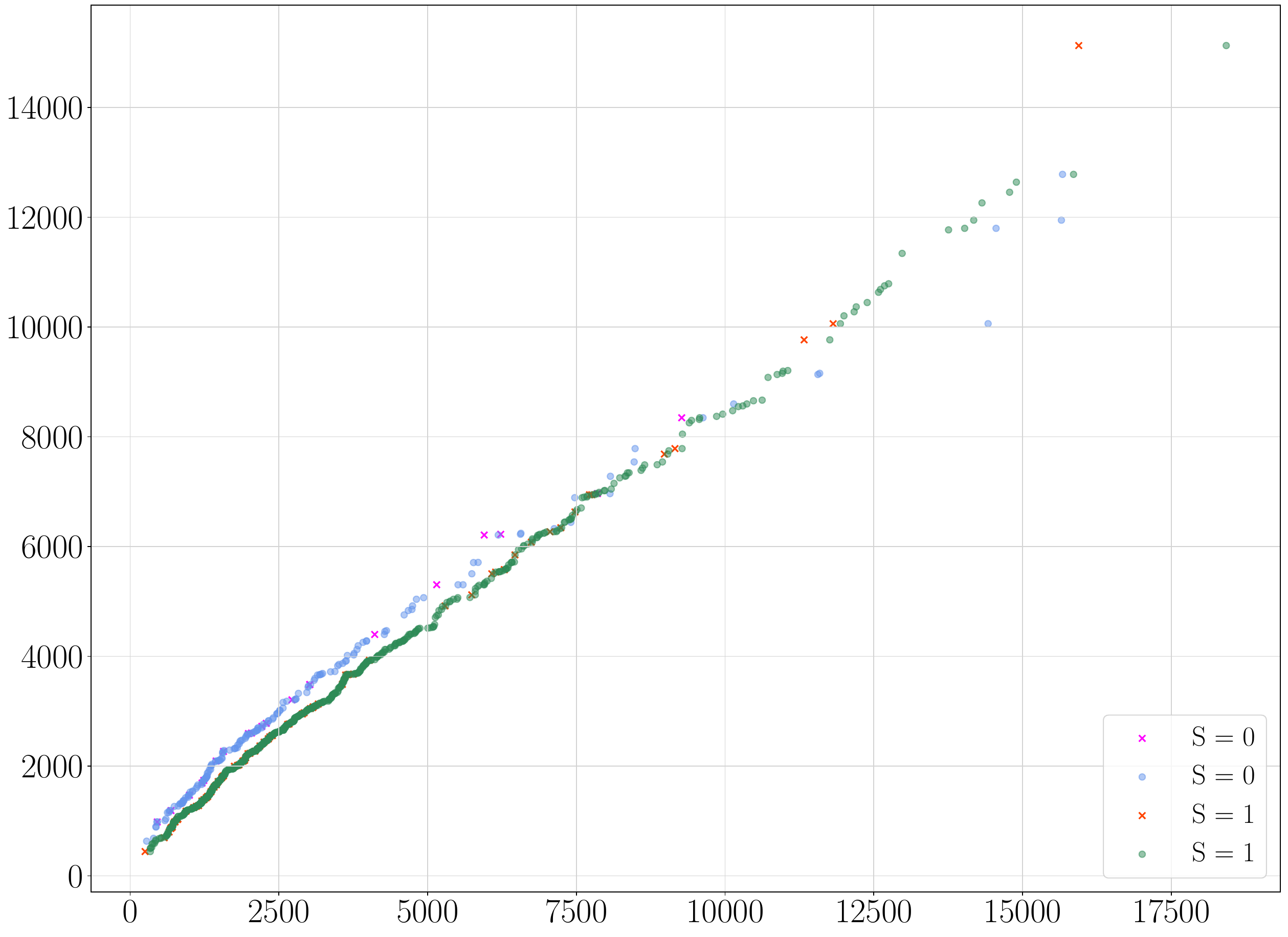}
        \caption{Original vs. repaired credit}
        \label{im:9a}
    \end{subfigure}
    \begin{subfigure}[b]{0.31\textwidth}
    \includegraphics[width=\textwidth]{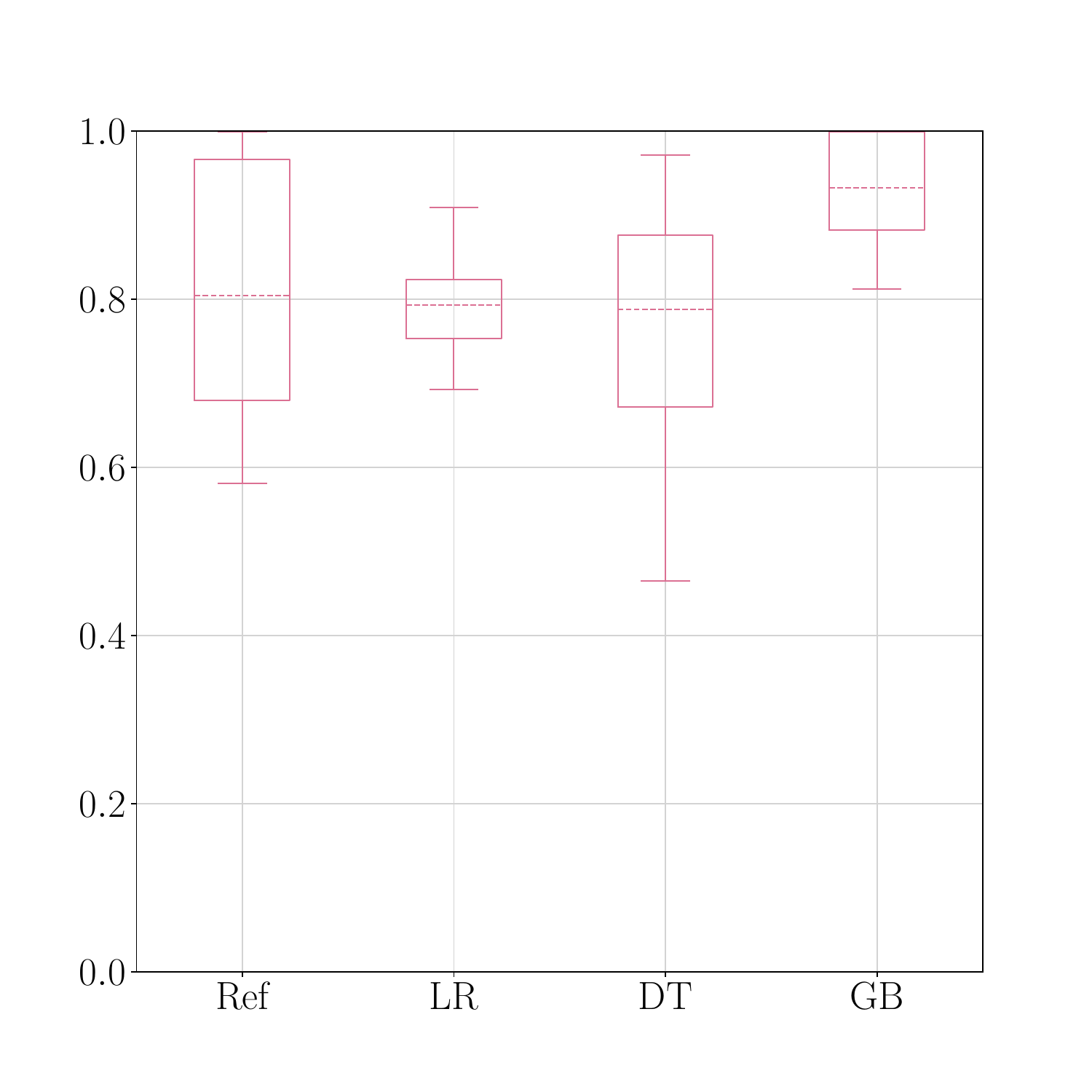}
    \caption{Disparate impact}
         	\label{im:9b}
   \end{subfigure}
    \begin{subfigure}[b]{0.31\textwidth}		\includegraphics[width=\textwidth]{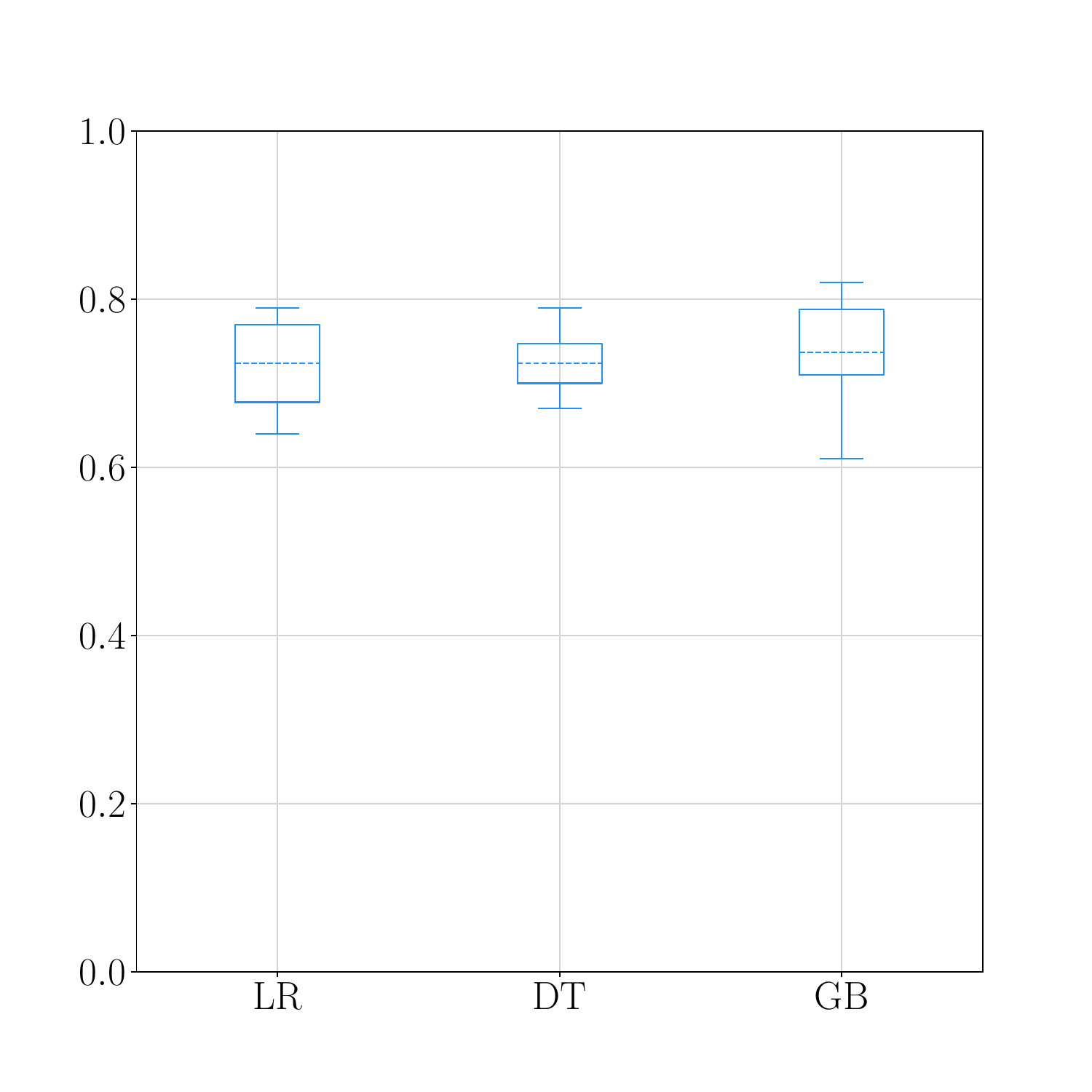}
         	\caption{Accuracy}
         	\label{im:9c}
     	\end{subfigure}

\begin{subfigure}[b]{0.36\textwidth}
\includegraphics[width=\textwidth]{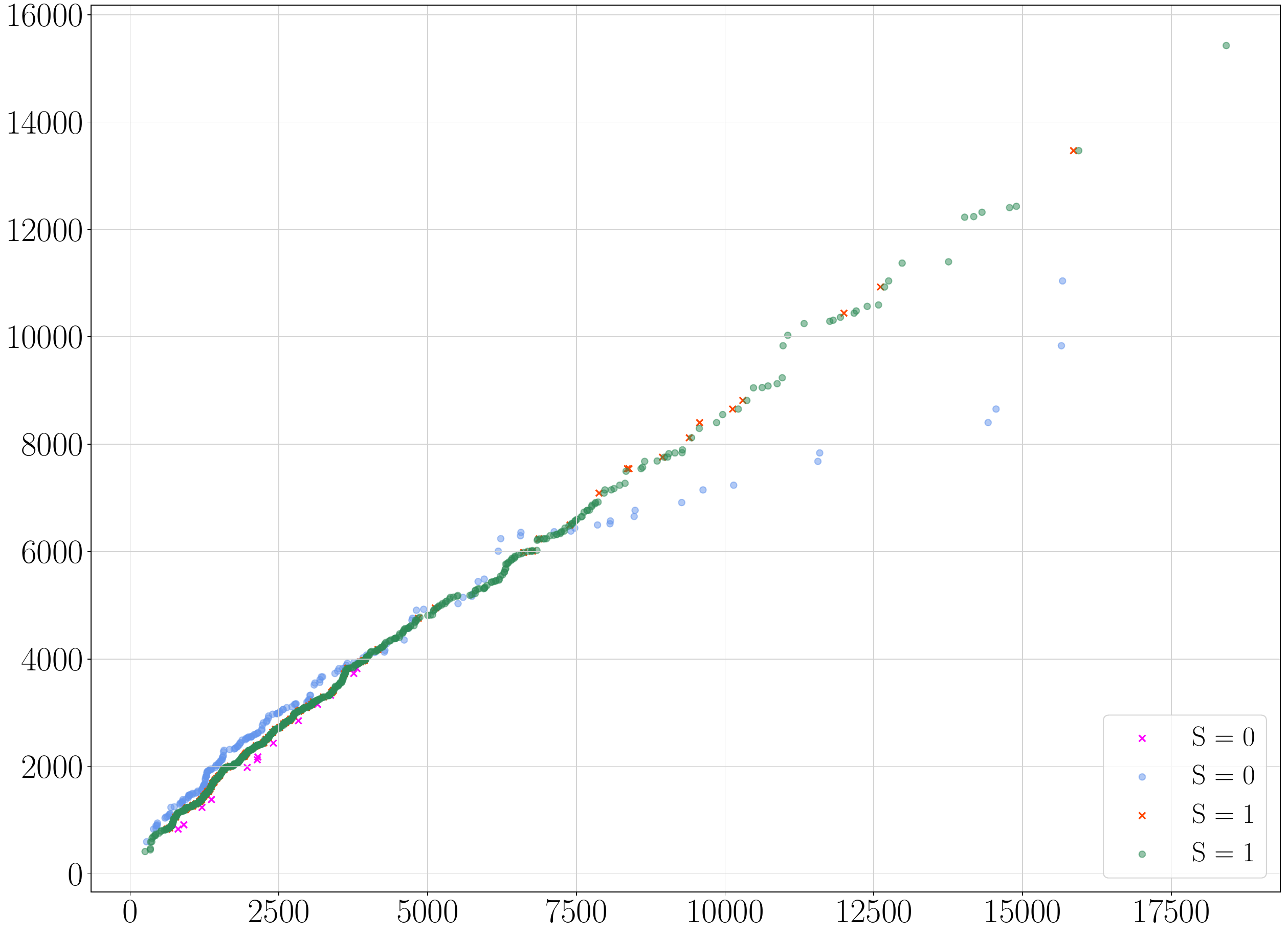}
\caption{Original vs. repaired credit}
         	\label{im:10a}
     	\end{subfigure}
     	\begin{subfigure}[b]{0.31\textwidth}
\includegraphics[width=\textwidth]{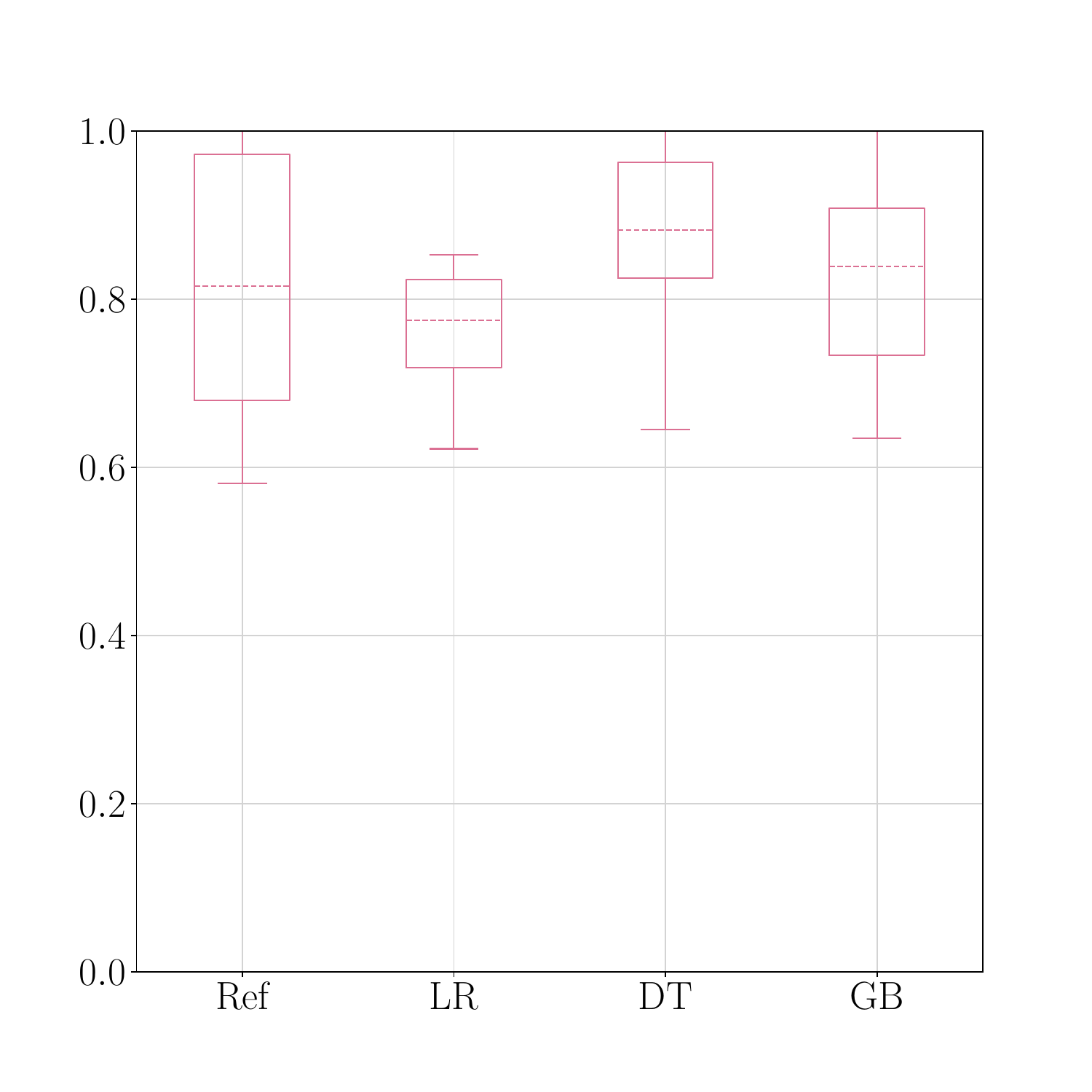}
 \caption{Disparate impact}
         	\label{im:10b}
     	\end{subfigure}    	\begin{subfigure}[b]{0.31\textwidth}	\includegraphics[width=\textwidth]{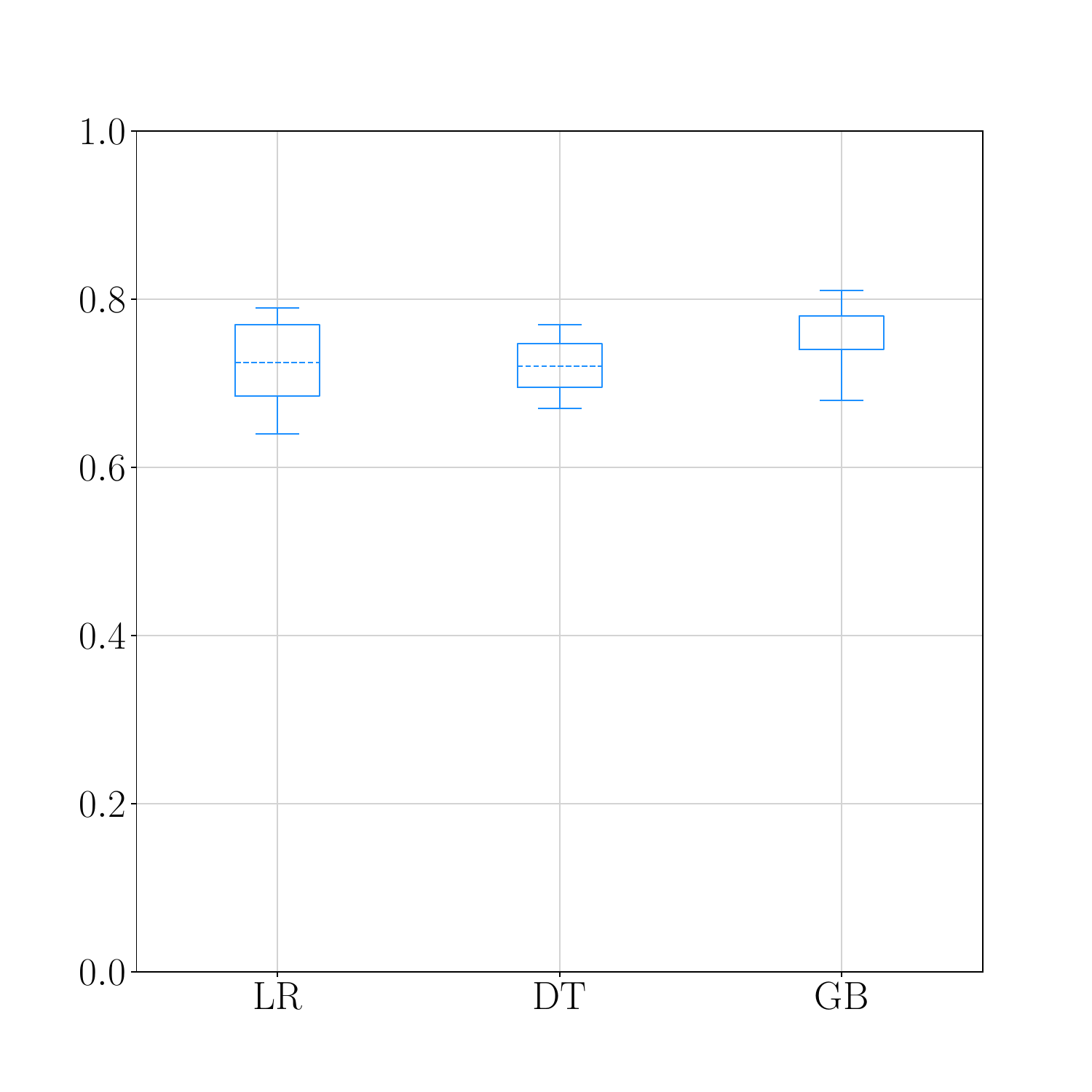}
        \caption{Accuracy}
         	\label{im:10b}
     	\end{subfigure}
        
    		\caption{Results of Experiment ($\mathcal{R}$2). First row: Option 1 - no regularization step. Second row: Option 2 - regularization step. Third row : Option 3 - hybrid version. In column (a) : Original vs. repaired credit amount for the training set (blue $S=0$, green $S=1$ points) and the test set (pink $S=0$, orange $S=1$, crosses). In column (b) : Disparate impact of classifiers LR, DT, GB. In column (c) : Accuracy of classifiers LR, DT, GB.}
    	\label{im:exp_real_data_option1}
	\end{figure}

\begin{itemize}

	\item[($\mathcal{R}$1)] \textit{Via \textbf{Classical Benchmark} assessment.} We conducted an initial experiment to quantify the extent to which the preprocessing bias mitigation techniques improved the fairness metrics in comparison to the metrics obtained by the predictive algorithms trained on the original dataset. As illustrated in the Figure \ref{im:real-data-benchmark}, the average ratio between the positive prediction for the different groups is close to $0.8$, which means that, according to the historical data, for every $5$ individuals with successful outcome ($Y = 1$) in the group $\text{age} > 25$, $4$ in the group $\text{age} \leq 25$ class have a successful outcome too. For this particular case study and using the default parameterizations for the models, we obtain lower ratio for the predictions with LR and GB on average. Finally, regarding precision, the model that performs better is the GB.

    \begin{figure}[h!]
     \begin{subfigure}[b]{0.4\textwidth}
     \centering
         \includegraphics[width=\textwidth]{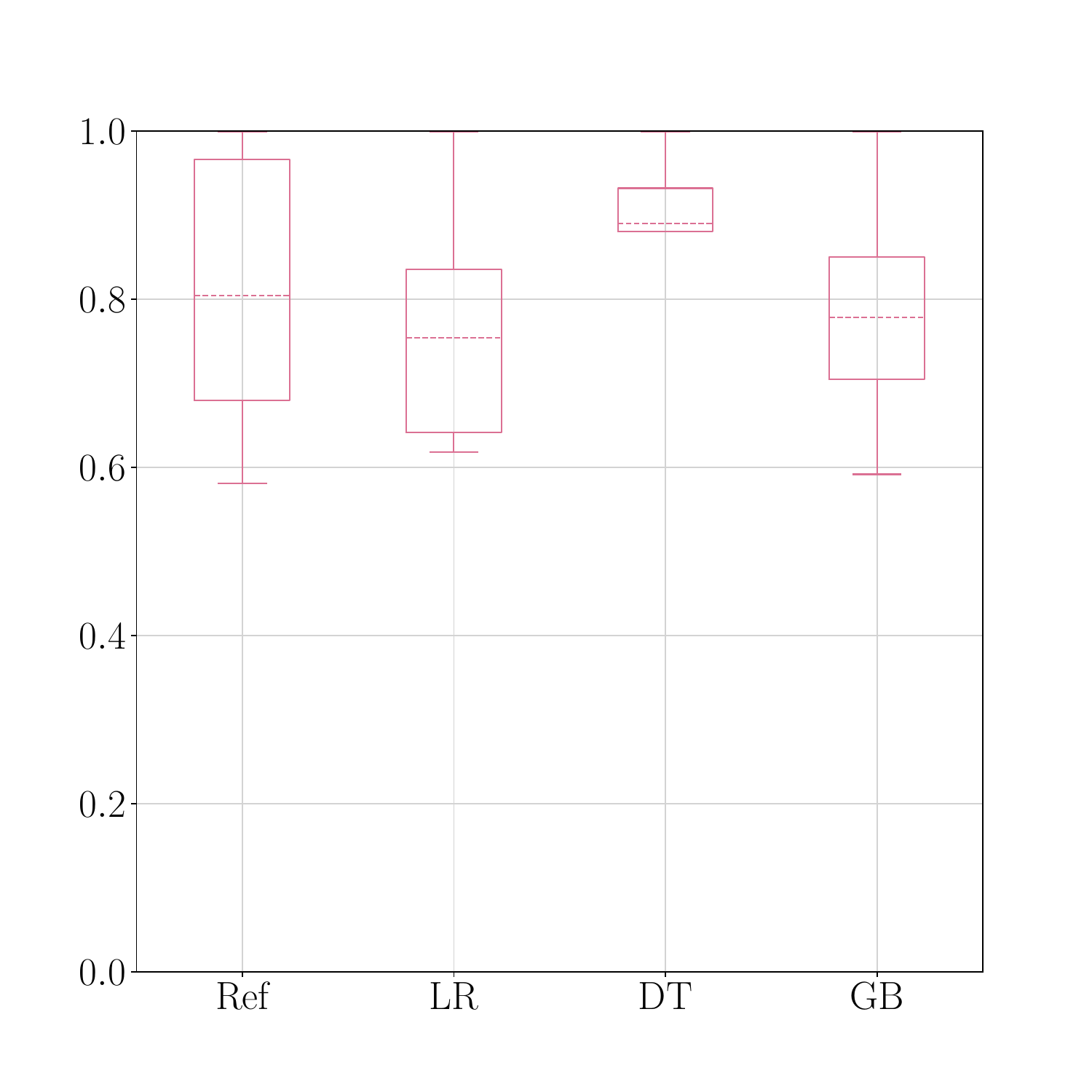}
         \caption{Disparate impact.}
         \label{im:rdba}
     \end{subfigure}
     \begin{subfigure}[b]{0.4\textwidth}
     \centering
         \includegraphics[width=\textwidth]{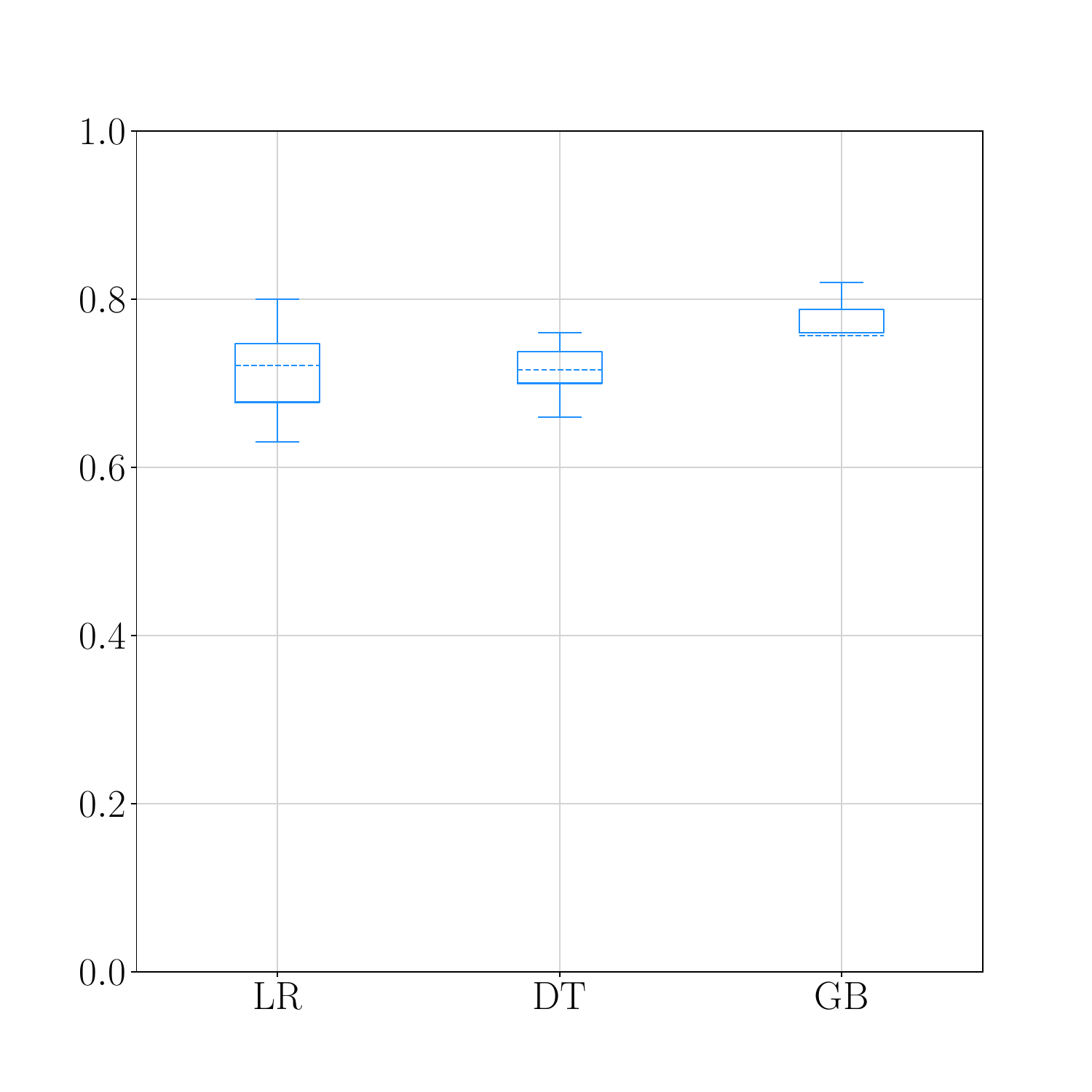}
         \caption{Accuracy.}
         \label{im:rdbb}
     \end{subfigure}
    \caption{Results of Experiment ($\mathcal{R}$1). Performance (a) and fairness (b) metrics across the three models under consideration without altering the dataset. The Ref values are the dataset ones.}
    \label{im:real-data-benchmark}
    \end{figure}
    
    \item[($\mathcal{R}$2)] \textit{Via \textbf{Extended Total Repair} assessment.} To correct for the bias detected in the benchmarking results, we proposed to apply the OT interpolation methodology within a set validation approach. In each iteration, the training data is processed using the total repair procedure, while the test data is handled using the extended total repair method, as detailed in Section \ref{sec:methodology}. We implemented the three possibilities for the interpolation step mentioned in Section \ref{sec:discussion_implementation}. However, for this particular case study, we note that the appropriate option would be the hybrid version (Option 3) due to the fact that for each fold, there are a high amount of samples where the variable of interest (credit amount) is less than $ \$ 6000$. Hence, for the values below such a threshold, we suggest not to add the Moreau-Yosida regularization in order to compute the repair value. On the contrary, for those above the threshold we encourage to construct a smooth interpolation function. 
    Such decisions, e.g. setting the threshold, aligns with each specific case study, depending on the respective data.
    \par
For each of the computational options, we show the results of the interpolation for one of the k-folds in the first column of Figure \ref{im:exp_real_data_option1}. 
%
The proposed method is applied using discrete OT maps for the training set (Section \ref{sec:total-repair}) and the continuous extension for the test data (Section \ref{sec:theo_extended}).
    The cyclically monotone graphs of $T_s$ and its interpolation $\Bar{T}_{s}$ are represented. Precisely, the training points and its repaired versions are drawn in: blue dots for $S = 0$ (minority group), and green dots for $S = 1$ (majority group). Moreover, the testing sets and their interpolated repaired version are drawn in: pink crosses for $S=0$, and purple crosses for $S=1$.
    
In addition, we empirically study the fairness-accuracy trade-off, for which results are shown in the second and third columns of Figure \ref{im:exp_real_data_option1}. Concerning the precision of the different binary classifiers employed, no significant differences emerge across the three interventions in preprocessing historical biased data. 
This observation leads us to infer that the precision of the repair values does not  exert a substantial influence on the reduction in the accuracy of the prediction. 
Conversely, the fairness metric DI exhibits variation in tandem with the complexity of classifiers and the precision of reparation in the test data. It is important to note that at least one of the trained algorithms achieves a significant increase in DI.
\par
As discussed above, these three alternatives for computing the interpolated OT map have appeared in the natural process of adapting our proposal from simulated data to real data. Indeed, in real scenarios, where the data to be transformed is not usually uniformly distributed across the feature space, this adaptation allows for improved performance of our methodology. In short, the recommended approach will depend on the specific characteristics of the data and the problem at hand. 
\end{itemize}

\section{Conclusions}
\label{sec:conclusions}
The use of some demographic variables during the learning process of an algorithm may lead to discriminatory outcomes against minority groups within the population, with respect to a sensitive or protected variable. In most cases, companies are not willing to give access to these algorithms or data. This paper  addresses fair classification problems, specifically focusing on  appropriately preprocessing biased data, from an OT theory approach.
In our study we measured such a bias through the disparate impact, a well-known  fairness metric based on the algorithm's predictions across the different groups.
\par
Previous work has established a statistical framework aimed at mitigating the value of the protected class in order to obtain a fair classification. This methodology relies on transporting in an optimal way the empirical distributions of the unprotected characteristics, conditioned to the protected variable, to their weighted Wasserstein barycenter \citep{pmlr-v97-gordaliza19a}.
Nevertheless, a remarkable drawback of this procedure emerges when there is a requirement to transport new data drawn from the same distributions. In such cases, it would become imperative to recompute the transport maps using the augmented dataset, leading to a computationally inefficient procedure. Alternatively, based on \cite{10.1214/20-AOS1996}, we developed a computationally efficient approach for transporting novel points 
to its corresponding version at the previously calculated Wasserstein barycenter. Our contribution is called the Extended Total Repair (ExTR).
\par
The efficient solution involves finding an smooth interpolation function under cyclically monotonicity. Such a property generalises to high dimensions the preservation of the total order on the real line, very suitable for fairness purposes.
This function is the subgradient of a convex function computed via the stochastic subgradient descent algorithm, whose smoothness relies on a specific parameter, determined by an exact algorithm. 
Specifically, the parameter is the minimum cycle mean of the associated graph of the dual formulation of the optimization problem to obtain the optimal smoothing value. The computation of such a value has been already proposed using Karps’ algorithm. In this paper, we improved this solution by providing an efficient algorithm inspired by a type of auction algorithm, which solves the assignment problem at less computational cost than the exact Hungarian algorithm.
\par
In this work there are two main issues we had to deal with. On the one hand the need to align the samples at the beginning of the process. On the other hand the  selection process aiming for  smoothest interpolation function.
\par
We carried out several experiments to evaluate the ExTR under different scenarios both with simulated and real data. Importantly, we note that the simulations have led us to conclude a practical adaptation of our methodology to obtain better results when applied to real data. The ExTR has proved to coherently transform data outside the support from which the original reparation is computed. In addition, the DI was improved at least for one of the three trained classifiers, while the accuracy remained unchanged.
\par
Finally, we would like to mention that the methodology is  generalizable to multi-class protected variables, since it only affects the computation of the Wasserstein barycenter. Therefore, we believe that it would be interesting as future work to test it in this context, for which the corresponding version of DI should be used.

 \subsubsection*{Acknowledgments}

The first author was supported by the Gobierno de Navarra through the MRR Investigo 2023 program under the grant 0011-4001-2023-000071.
The second author was supported in part by the Spanish Ministry of Science and Innovation under the grant PID2021-128314NB-I00.
The third author was supported by Ministerio de Ciencia e Innovación PID2020-113443RB-C21.

 The authors would like to thank Eustasio del Barrio Tellado for helpful discussions.


\newpage

\appendix

\section{Background}
\label{app:sec:background}
This section is dedicated to briefly introduce certain notions on some mathematical theory and problems that are required for our proposal.
First, a foundational overview of optimal transport theory is provided as the basis of the data transformation methodology. 
Specially the total repair procedure hinges on the Brenier's existence theorem. Afterwards, we describe the minimum mean cycle problem, which is related to the calculus of some parameter needed for constructing the interpolation function of the methodology. Finally, we describe the assignment problem that will be important in the efficient computation of the previously mentioned value.

\subsection{Optimal transport}
\label{app:sec:background:OT}
\par
Optimal transport (OT) theory is a branch of mathematics that deals with the problem of finding the most efficient way to transport mass from one distribution to another. It was originally introduced by the French mathematician Gaspard Monge in 1781 in the context of military logistics in a very famous paper \textit{M\'emoire sur la Th\'eorie des D\'eblais et des Remblais}, and later refined by Leonid Kantorovich in the 1940s as a tool for economics.
In recent years, there has been a surge of interest in OT theory due to its potential for solving challenging problems in a wide range of fields. In particular, the development of numerical algorithms and efficient computational methods has made it possible to apply OT to large-scale problems, opening up new opportunities for research and applications in the field of machine learning (ML). Domain adaptation \citep{courty2016optimal} , transfer learning \citep{gayraud2017optimal}, natural language processing (NLP) \citep{xu2018distilled}, clustering \citep{ho2017multilevel} or fair learning \citep{pmlr-v97-gordaliza19a, jiang2020wasserstein} are some examples of ML problems that have been addressed through OT.
\par
For a domain $\Omega \subseteq \mathbb{R}^{d}$, let $\mathcal{P}(\Omega)$ be the space of Borel probability measures with support contained in $\Omega$ and $\mathcal{P}_{ac}(\Omega)$ be those with densities.
In this context, the pushforward of a measure $P$ by a measurable function $T: \mathbb{R}^{d} \rightarrow \mathbb{R}^{d}$ is denoted as $T\#P$, and satisfies $(T\#P)(A) = P(T^{-1}(A))$ for every measurable set $A$. 
The objective of OT is to find a transport map between two probability measures $P$ and $Q$ that minimizes the average cost of transportation between two measures given by a Borel function $c: P \times Q \rightarrow \mathbb{R} \cup \{ \infty \}$. 
One way to express Monge's problem \citep{monge1781mémoire} in a simple and probabilistic form is as follows: 
\begin{equation}
    \label{eq:3}
    \inf_{T \in \mathcal{T}(P, Q)} \int_{\mathbb{R}^{d}} c(\boldsymbol{x}, T(\boldsymbol{x})) dP(\boldsymbol{x}),
\end{equation}
where $\mathcal{T}(P, Q) = \{T: \Omega \rightarrow \Omega \text{  }|\text{  } T\# P = P \circ T^{-1} = Q\}$ is the set of transport maps from $P$ to $Q$, two measures in $\mathcal{P}^{2}(\mathbb{R}^{d})$ (space of probability measured having a finite second moment on $\mathbb{R}^{d}$). 
A map achieving the infimum in \eqref{eq:3} is called an OT map of $P$ to $Q$. 
The existence and uniqueness of an optimal map between $P$ and $Q$ is proven when the first measure is absolutely continuous. This fundamental result is due to \cite{https://doi.org/10.1002/cpa.3160440402}. Essentially, the Brenier' Theorem asserts that as long as the source measure has a density, a unique OT map exists and it is the gradient of a convex function, see Theorem 1.22 and Theorem 1.40 of \cite{santambrogio2016optimal}.
\par
The OT problem is a broader concern that can be stated in general metric spaces, such as Polish probability spaces, as detailed in  Theorem 4.1 of the reference \cite{villani2008optimal}.
However, for the context of our paper, we restrict to the metric space $\mathbb{R}^{d}$ where the cost function assumes the form of the Euclidean cost $c(\boldsymbol{x}, \boldsymbol{\tilde{x}}) = ||\boldsymbol{x} - \boldsymbol{\tilde{x}}||_{2}^{p}$ with $p \geq 1$.
For this particular case, the value of the optimization problem Eq. \eqref{eq:3} is then the Wasserstein distance of order $p$. 
Notably, in the instance where $p = 2$, the squared $2$ - Wasserstein distance between $P$ and $Q$ is defined as:
\begin{equation}
    \label{eq:4}
    \mathcal{W}_{2}^{2} (P, Q) = \min_{\pi \in \Pi(P,Q)} \int \Vert \boldsymbol{x} - \boldsymbol{\tilde{x}} \Vert ^{2}_{2} d\pi(\boldsymbol{x},\boldsymbol{\tilde{x}}),
\end{equation}
where $\Pi(P,Q)$ is the set of probability measures on $\mathrm{R}^{d} \times \mathrm{R}^{d}$ with marginals $P$ and $Q$. 
\par
The Wasserstein metric $\mathcal{W}_{2}^{2}$ was initially introduced by \cite{KR:58} to quantify distances between probability distributions.
According to the findings of \cite{villani2003topics}, for $P, Q \in \mathcal{W}_{r} (\mathrm{R}^{d})$ (Wasserstein space: space of probability measures over $\mathrm{R}^{d}$ with finite $r$-order) and $1 \leq r < \infty$, the infimum in Eq. \eqref{eq:4} is reached for some measure $\pi \in \Pi(P,Q)$ \citep{https://doi.org/10.1002/cpa.3160440402}.

In the realm of fair learning, the squared $2$- Wasserstein metric $\mathcal{W}_{2}^{2}$ has proven to be a suitable way to measure the similarities between distributions. In our study, it compares probability distributions between a variable $X$ and another $\tilde{X}$, obtained from a non-uniform transformation in $X$, whether it is random or deterministic. We refer the reader to the text \cite{villani2008optimal} for more details on OT theory in general.
\subsubsection{Wasserstein barycenters}
Furthermore, we will delve into the theoretical concept of the mean of a collection of probability measures.
This generalization to metric spaces, not necessarily linear, is the $2$ - Wasserstein barycenter, which is unique if the measures are absolutely continuous. For our work, we will optimally transport the conditional distributions of the data with respect to the group ($S = 0$ or $S = 1$) into this "mean" law, properly defined as following:

\begin{definition}
Given probability measures $(\mu_{j})_{1 \leq j \leq J}$ with finite second moment and positive, summing up to 1, weights $(w_{j})_{1 \leq j \leq J}$, the Wasserstein barycenter is a minimizer of 
\begin{equation*}
    \nu \mapsto \sum_{j=1}^{J} w_{j} \mathcal{W}_{2}^{2} (\nu, \mu_{j}).
\end{equation*}
\end{definition}

We refer to \cite{doi:10.1137/100805741} for further details on the theory of Wasserstein barycenters and to \cite{10.1561/2200000073},  \cite{pmlr-v32-cuturi14} or \cite{boissard2015distribution} for computational methods. 
In a broader context, the Wasserstein barycenter proves to be a meaningful representative of the average prototype within a set of distributions; in addition to have naturally inherit the ability of OTation to capture geometric properties of the data.

\subsection{Minimum cycle mean in a digraph}
\label{app:sec:background:mcm}
\par
The minimum mean cycle problem can be considered as a particular case of the minimum cost-to-time ratio cycle problem - finding a directed cycle for which the sum of the costs divided by the sum of the transit times associated to the edges is minimum - \citep{Lawler1972}. 
The issue has previously undergone examination through works as \cite{CHATURVEDI201721}. Moreover, an approximation algorithm was employed in an attempt to address it as \cite{CHATTERJEE2014104}.
\par
The formal problem formulation is as follows. Consider a finite directed graph with a set $V$ of $n$ vertices denoted as $G = (V, E)$, where $E$ represents the set of edges (ordered pairs of vertices), often called arcs. 
Within this context, let $f$ be a function from $E$ into the real numbers, associating with each edge $e \in E$, a weight $f(e)$. 
For any given sequence of edges denoted as $\sigma = (e_{1}, e_{2}, ..., e_{p})$, we define its weight $w(\sigma)$ as $\sum_{i=1}^{p} f(e)$. Furthermore, we introduce the concept of mean weight $m(\sigma)$, characterizing the average weight as the cumulative weight divided by the length of the sequence $w(\sigma)/p$. 
A directed cycle $W_{d}$ in a directed graph is a non-empty directed trail in which only the first and last vertices are equal. 
The minimum mean cycle problem consists in identifying a cycle whose mean cost is minimum. However, the primary interest lies in determining the minimum cycle mean designed as $\lambda^{\star}$ and defined as:
\begin{equation*}
    \lambda^{\star} = \min_{\mathcal{W}_{d}} m(W_{d}),
\end{equation*}
where $W_{d}$ ranges over all directed cycles in $G$, denoted as $\mathcal{W}_{d}$.

One method involves employing the binary search algorithm. In this approach, each iteration entails solving a shortest path problem with arbitrary lengths \citep{lawlere:76}.
Alternatively, \cite{karp1978characterization} provides both a characterization of $\lambda^{\star}$, as well as an algorithm for its computation, with a time complexity of $\mathcal{O}(n |E|)$, where $|E|$ denotes the cardinal of E.

\subsection{The linear assignment problem}
\label{app:sec:background:AP}
\par
The problem of weighted matching in a bipartite graph, often referred to as the assignment problem (AP), holds a prominent place in the field of combinatorial optimization due to its extensive research and applications in various practical settings. 
The literature dedicated to the problem is extensive and we refer to \cite{ahuja1989chapter} for a detailed survey of algorithmic problem-solving methods.
\par
A bipartite graph is a graph where the $n$ vertices can be divided into two disjoint sets such that all edges connect a vertex in one set to a vertex in another set. There are no edges between vertices in the disjoint sets. 
For formulating the assignment problem, let us consider a set $N_{1}$ denoting persons who wish to assign among themselves $N_{2}$ objects, where we assume, without any loss of generality, that $|N_{1}| = |N_{2}|$.
There exists a collection of pairs $A \subseteq N_{1} \times N_{2}$ which represents potential assignments of individuals to objects. 
Within this collection, each element $(i,j) \in A$ possesses an associated integer cost $c_{i,j}$. 
The objective entails the assignment of each individual to exactly one object while minimizing the overall assignment cost.
This problem can be precisely formulated within the context of a linear program, as follows:
\begin{equation*}
\begin{cases}
    \text{Minimize} \sum_{(i,j) \in A} c_{i,j}x_{i,j} \text{    subject to:}\\
    \sum_{\{j:(i,j) \in A\}} x_{i,j} = 1, \forall i  \in N_{1}, \\
    \sum_{\{i:(i,j) \in A\}} x_{i,j} = 1, \forall j  \in N_{2}, \\
    x_{i,j} \in \{0,1\}, \quad \forall (i,j) \in A.\\
\end{cases}       
\end{equation*}
\par
It is worth noting that the assignment problem falls under the broader category of the transportation problem, which, in turn, is a special case of the minimum cost flow problem. 
Different methods for solving this problem range from primal-dual and successive shortest paths, cost parametric, iterative, easing, relying on signatures, and primal approaches, among others (see \cite{10.1007/978-3-642-77489-8_5} for a survey).
Among the various techniques to tackle the assignment problem, the successive shortest path algorithm stands out as a popular choice. 
To apply this algorithm, we first need to transform the assignment problem into a minimum cost flow problem and solve it over an extended graph by adding a source node $s$, a sink node $t$ and arcs (directed edges) $(s, i)$ for all $i \in N_{1}$ and $(j, t)$ for all $j \in N_{2}$. These arcs will have zero cost and unit capacity.
The algorithm in $n$ applications successively obtains a shortest path from $s$ to $t$ with respect to the linear program reduced costs of the extended graph $\tilde{c}_{i,j} = c_{i,j} - \pi(i) + \pi(j)$, updates the node potentials $\pi$ (dual variable with the mass balance constraint of a node), and augment one unit of flow along the shortest path. 
\par
Alternatively, another approach to address the assignment problem is the auction algorithm, initially proposed by \cite{Bertsekas_auction}. 
This method operates on a different conceptual basis, assigning the objects to individuals through auctions with the aim of reaching a state of economic equilibrium. 
\par
In a creative fusion of these two concepts, \cite{Orlin1992} introduced a hybrid version of the auction and successive shortest path algorithms. 
This innovative approach runs in time $\mathcal{O}(\sqrt{n} |A| \log (n C_{bound}))$, which substantially improves the running times obtained by using either technique alone. In this context $C_{bound} = \max \{ c_{i,j} \} + 1$.

\section{Additional details and results}

\subsection{Additional results and plots}
\label{app-add-plots}
\begin{figure}[h!]
    \centering
    \includegraphics[width=.45\textwidth]{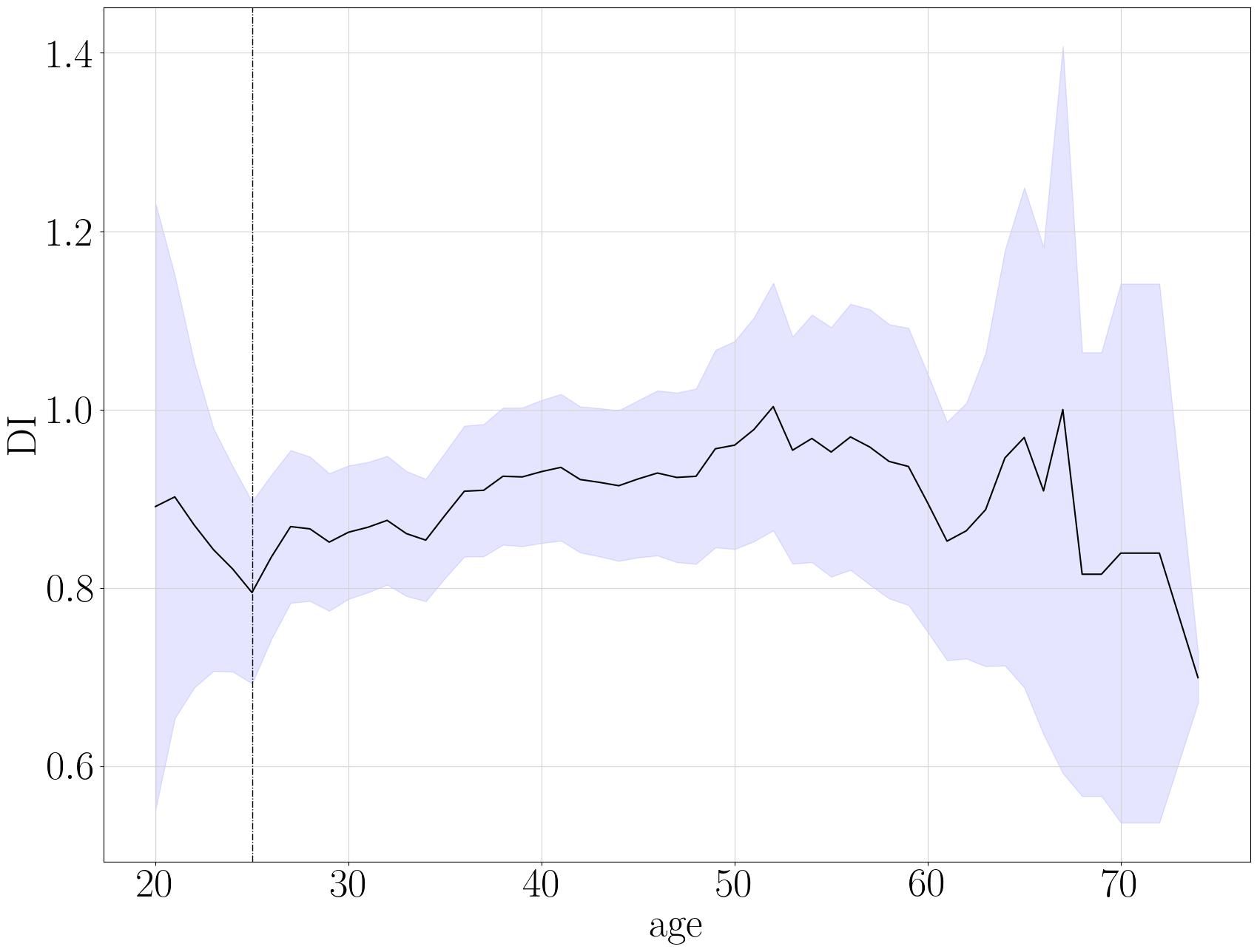}
    \caption{$95 \%$ confidence interval of Disparate impact. The x-axis represents the threshold used to binarise age as protected attribute.}
    \label{im:threshold}
\end{figure}

Figure \ref{im:threshold} represents the $95 \%$ confidence interval for the disparate impact for different age thresholds ($x$-axis). 

Table \ref{table:comp_costs} shows the power of the proposed procedure to compute repair values for new data.


\bibliographystyle{unsrtnat} 
\bibliography{bibliography.bib}       



 \begin{table}[b]
    \caption{Experiment $\mathcal{E}$1: Fairness (disparate impact) and performance (accuracy) metrics measured in the original dataset and in the modified dataset. The results are the average values of the $10$-fold cross validation.}
      \centering
      \begin{tabular}{rllll}
        \hline
      Configuration &  Data & Model & Disparate impact & Accuracy \\
        \hline
       \multirow{6}{5em}{$\mathcal{E}$1-A} & \multirow{3}{10em}{Original dataset} & LR & $0.449 \pm 0.089$ & $0.822 \pm 0.056$\\
        & & DT & $0.56 \pm 0.14$ & $0.847 \pm 0.05$\\
       & & GB & $0.558 \pm 0.156$ & $0.895 \pm 0.033$\\
        \cline{2-5}
       & \multirow{3}{10em}{Transformed dataset} & LR & $0.758 \pm 0.07$ & $0.753 \pm 0.056$\\
      &  & DT & $0.74 \pm 0.168$ & $0.805 \pm 0.049$\\
     &   & GB & $0.692 \pm 0.189$ & $0.878 \pm 0.033$\\
        \hline
      \multirow{6}{5em}{$\mathcal{E}$1-B} &  \multirow{3}{10em}{Original dataset} & LR & $0.339\pm 0.093$ & $0.824 \pm 0.033$\\
      &   & DT & $0.481 \pm 0.147$ & $0.834 \pm 0.047$\\
      &   & GB & $0.413 \pm 0.144$ & $0.914 \pm 0.041$\\
        \cline{2-5}
      &   \multirow{3}{10em}{Transformed dataset} & LR & $ 0.721 \pm 0.081$ & $ 0.77 \pm 0.059 $ \\
    &     & DT & $0.719 \pm 0.114$ & $0.842 \pm 0.032$ \\
     &    & GB & $0.662 \pm 0.115$ & $0.852 \pm 0.029$ \\
        \hline
      \end{tabular}
      \label{table:exp2}
    \end{table}

\begin{table}
 \caption{Distribution of the credit risk level (class label) with respect to  gender (left) and  age (right).}
    \centering
    \begin{tabular}{c|cc||cc|c}
   & Female & Male & Young & Senior & Total\\ \hline
  Credit       &  109&191&80&220 & 300\\
  No credit       & 201&499&110&590 & 700
    \end{tabular}
    \label{tab:6}
\end{table}

    \begin{table}[h!]
  \caption{Bias measured in the original dataset. The CI with $95\%$ confidence were computed using the method described by \cite{besse2018confidence}.}
  \centering
  \begin{tabular}{lll}
    \hline
    Sensitive feature & Disparate impact & Confidence Interval ($\alpha = 0.05$)\\
    \hline
    Gender \{male vs female\} & 0.897 & [0.812, 0.981]\\
    Age \{$> 25$ vs $\leq 25$ \} & 0.795 & [0.693, 0.897]\\
    \hline
  \end{tabular}

  \label{table:di}
\end{table}

\begin{table}[h!]
 \caption{Average and standard deviation on the 10 fold CV of the disparate impact and accuracy measures when considering S= 'age' the protected attribute.}
      \centering
      \begin{tabular}{llll}
        \hline
        Experiment & Model & Disparate impact & Accuracy\\
        \hline
        \hline
        \multirow{3}{8em}{($\mathcal{R}$1) Benchmark} & LR & $0.754 \pm 0.119$ & $0.721 \pm 0.056$\\
        & DT & $0.89 \pm 0.082$ & $0.716 \pm 0.03$\\
        & GB & $0.778 \pm 0.111$ & $0.757 \pm 0.05$\\
        \hline
        \multirow{3}{8em}{($\mathcal{R}$2) ExTR - Option 1} & LR & $0.741 \pm 0.108$ & $0.729 \pm 0.051$ \\
        & DT & $0.893 \pm 0.061$ & $0.722 \pm 0.038$\\
        & GB & $0.806 \pm 0.103$ & $0.747 \pm 0.052$\\
        \hline
        \multirow{3}{8em}{($\mathcal{R}$2) ExTR - Option 2} & LR & $0.793 \pm 0.1$ & $0.724 \pm 0.054$ \\
        & DT & $0.788 \pm 0.155$ & $0.724 \pm 0.038$\\
        & GB & $0.932 \pm 0.102$ & $0.737 \pm 0.06$\\
        \hline
        \multirow{3}{8em}{($\mathcal{R}$2) ExTR - Option 3} & LR & $0.775 \pm 0.101$ & $0.725 \pm 0.054$ \\
        & DT & $0.88 \pm 0.11$ & $0.722 \pm 0.033$ \\
        & GB & $0.839 \pm 0.139$ & $0.74 \pm 0.056$\\ 
        \hline
      \end{tabular}
      \label{table:4}
\end{table}

\begin{table}[h!]
 \caption{Computational costs measured in seconds for one dimensional simulated data. DOTP refers to \textit{discrete OT plan}. Notation: $k_{s}$ refers to the number of new samples of each class $s \in \{0,1\}$.}
  \centering
  \begin{tabular}{cccccc}
    \hline
    \multicolumn{4}{c}{Experimental setup} & \multicolumn{2}{c}{Execution time}           \\
    \cmidrule(r){1-4} \cmidrule(r){5-6}
    $n_{0}$  & $n_{1}$ & $k_{0}$ & $k_{1}$ & Recompute $DOTP^{\star}$ & Methodology proposed \\
     \hline
     \hline
    20 & 20 & 1 & 1 & 0.0625 & 0.0157 \\
    100 & 100 & 1 & 1 & 0.9067  & 0.0189 \\
    100 & 100 & 20 & 20 & 1.3752 & 0.1404\\
    200 & 200 & 1 & 1 & 4.6330 &  0.0157\\
    200 & 200 & 40 & 40 & 7.4161 & 0.1718\\
     \hline
  \end{tabular}
 
  \label{table:comp_costs}
\end{table}

\end{document}